\documentclass[12pt]{article}

\ifx\pdfoutput\undefined
\usepackage[dvips,bookmarks]{hyperref}
\else
\usepackage{hyperref}
\fi
\hypersetup{colorlinks=false,bookmarksopen,bookmarksnumbered,citecolor=blue,
   pdfstartview=FitH}

\usepackage[pdftex]{graphicx}	
\usepackage{latexsym}

\usepackage{amssymb,amsfonts,amsmath}
\usepackage{graphicx} 
\usepackage{indentfirst}

 \usepackage{bbm}

\parskip=\medskipamount

\newcommand {\cC}{{\cal C}}
\newcommand {\cD}{{\cal D}}
\newcommand {\cE}{{\cal E}}

\newcommand {\cJ}{{\cal J}}
\newcommand {\cK}{{\cal K}}
\newcommand {\cL}{{\cal L}}
\newcommand {\cM}{{\cal M}}
\newcommand {\cN}{{\cal N}}
\newcommand {\cO}{{\cal O}}

\newcommand {\cS}{{\cal S}}
\newcommand {\cT}{{\cal T}}
\newcommand {\cU}{{\cal U}}

\newcommand {\cX}{{\cal X}}

\newcommand{\bL}{{\bf L}}

\newcommand{\bR}{{\bf R}}

\def\a{\alpha}
\def \bi{\bibitem}

\def\b{\beta}
\def\c{\chi}
\def\d{\delta}

\def\g{\gamma}

\def\l{\lambda}
\def\m{\mu}
\def\n{\nu}
\def\o{\omega}
\def\p{\pi}
\def\q{\theta}
\def\r{\rho}
\def\s{\sigma}
\def\t{\tau}

\def\x{\xi}
\def\z{\zeta}
\def\D{\Delta}
\def\F{\Phi}

\def\L{\Lambda}
\def\O{\Omega}

\def\Q{\Theta}
\def\S{\Sigma}
\def\U{\Upsilon}

\def\tr{{\rm tr}}
\def\rd{{\rm d}}
\def\ri{{\rm i}}
\def\re{{\rm e}}

\newcommand{\ve}{\varepsilon}                            %new
\newcommand{\pa}{\partial}                           %new
\newcommand{\hf}{\frac12}
\newcommand{\vf}{\varphi}
\newcommand{\be}{\begin{equation}}
\newcommand{\ee}{\end{equation}}
\newcommand{\bea}{\begin{eqnarray}}
\newcommand{\eea}{\end{eqnarray}}
\newcommand{\non}{\nonumber}
\newcommand{\1}{{\underline{1}}}
\newcommand{\2}{{\underline{2}}}

\def\dt#1{{\buildrel {\hbox{\LARGE .}} \over {#1}}}    % dot-over for sp/sb

\newcommand{\bm}[1]{\mbox{\boldmath$#1$}}

\def\double #1{#1{\hbox{\kern-2pt $#1$}}}

\newcommand{\qb}{{\bar{\theta}}}

\newcommand{\OI}{{\overline{I}}}
\newcommand{\OJ}{{\overline{J}}}
\newcommand{\OK}{{\overline{K}}}
\newcommand{\OL}{{\overline{L}}}

\newcommand{\OP}{{\overline{P}}}
\newcommand{\OQ}{{\overline{Q}}}

\newcommand{\UI}{{\underline{I}}}
\newcommand{\UJ}{{\underline{J}}}
\newcommand{\UK}{{\underline{K}}}
\newcommand{\UL}{{\underline{L}}}

\newcommand{\UQ}{{\underline{Q}}}
\newcommand{\UP}{{\underline{P}}}

\newcommand{\bfD}{{\bf D}}
\newcommand{\bfDB}{{\bar{\bfD}}}

\newif\ifdtup

\def\de{{\nabla}}                                         % del
\def\deb{{\bar{\de}}}

\newcommand{\bsubeq}{\begin{subequations}}
\newcommand{\esubeq}{\end{subequations}}
\newcommand{\bai}{{\bar i}}
\newcommand{\baj}{{\bar j}}
\newcommand{\bak}{{\bar k}}
\newcommand{\bal}{{\bar l}}

\newcommand{\rL}{{\rm L}}
\newcommand{\rR}{{\rm R}}

\newcommand{\mub}{{{\bar{\mu}}}}

\begin{document}
%%%%%%%%%%%%%%%%
%%%%%%%%%%%%%%%%
\begin{titlepage}
\begin{flushright}
UUITP-11/12\\
May, 2012\\
\end{flushright}

\begin{center}
{\Large \bf 
Three-dimensional $\bm{(p,q)}$ AdS superspaces \\ and matter couplings
}\\ 
\end{center}

\begin{center}

{\bf
Sergei M. Kuzenko\footnote{Email: sergei.kuzenko@uwa.edu.au}${}^{a}$,
Ulf Lindstr\"om\footnote{Email: ulf.lindstrom@physics.uu.se}${}^{b}$, 
Gabriele Tartaglino-Mazzucchelli\footnote{Email: gabriele.tartaglino-mazzucchelli@physics.uu.se.
The  address starting June 2012: School of Physics M013, UWA,
35 Stirling Highway, Crawley W.A. 6009, Australia.}${}^{ab}$
} \\
\vspace{5mm}

\footnotesize{
${}^{a}${\it School of Physics M013, The University of Western Australia\\
35 Stirling Highway, Crawley W.A. 6009, Australia}}  
~\\
\vspace{2mm}

\footnotesize{
${}^{b}${\it Theoretical Physics, Department of Physics and Astronomy,
Uppsala University \\ 
Box 516, SE-751 20 Uppsala, Sweden}
}
\vspace{2mm}

\end{center}

\begin{abstract}
\baselineskip=14pt
We introduce $\cN$-extended  $(p,q)$ AdS superspaces in three space-time dimensions, 
with $p+q=\cN$ and $p\geq q$, and analyse their geometry.
We show that all $(p,q)$ AdS superspaces with $X^{IJKL}=0$ are conformally flat. 
Nonlinear $\s$-models with $(p,q)$ AdS supersymmetry exist for $p+q\leq 4$
(for $\cN>4$ the target space geometries are highly restricted). 
Here we concentrate on studying off-shell $\cN=3$ supersymmetric $\s$-models in $\rm AdS_3$. 
For each of the cases (3,0) and (2,1), we give three different realisations of the supersymmetric action.   
We show that (3,0) AdS supersymmetry requires the $\s$-model to be superconformal, and hence 
the corresponding target space is a hyperk\"ahler cone. 
In the case of (2,1) AdS supersymmetry, the $\s$-model target space must be 
a non-compact hyperk\"ahler manifold 
endowed with a Killing vector field which generates an
SO(2) group of rotations of the two-sphere of complex structures.
\end{abstract}

\vfill

\vfill
\end{titlepage}

\newpage
\renewcommand{\thefootnote}{\arabic{footnote}}
\setcounter{footnote}{0}

\tableofcontents{}
\vspace{1cm}
\bigskip\hrule

%%%%%%%%%%%%%%%%%%%%%%%%%%%%%%%%%%%%%%%%%%%%%%%%
%%%%%%%%%%%%%%%%%%%%%%%%%%%%%%%%%%%%%%%%%%%%%%%%
%%%%%%%%%%%%%%%%%%%%%%%%%%%%%%%%%%%%%%%%%%%%%%%%

\section{Introduction}
\setcounter{equation}{0}

In three space-time dimensions (3D), the anti-de Sitter (AdS)  group is reducible, 
$$\rm SO(2,2) \cong \Big( SL(2, {\mathbb R}) \times SL( 2, {\mathbb R}) \Big)/{\mathbb Z}_2~,$$ 
and so are its supersymmetric extensions,  
${\rm OSp} (p|2; {\mathbb R} ) \times  {\rm OSp} (q|2; {\mathbb R} )$.
This implies that  $\cN$-extended AdS supergravity exists in several incarnations  \cite{AT}. 
These are known as the  $(p,q)$ AdS supergravity theories\footnote{One should not confuse 
$(p,q)$ AdS supersymmetry 
in three dimensions with $(p,q)$ Poincar\'e supersymmetry in two dimensions \cite{Hull:1985v}.}   
where the  non-negative integers $p \geq q$ are such that $\cN=p+q$.   
For arbitrary values of $p$ and $q$ allowed, 
the pure  $(p,q)$ AdS supergravity was constructed in \cite {AT}
as a Chern-Simons theory with the gauge group
 ${\rm OSp} (p|2; {\mathbb R} ) \times  {\rm OSp} (q|2; {\mathbb R} )$.
 Similar ideas can readily be used, e.g., to construct 3D higher-spin  $(p,q)$ 
 $\rm AdS$ supergravity  \cite{Henneaux:2012ny}. 
However, this Chern-Simons construction appears to become less powerful when it comes to 
coupling 
AdS supergravity to supersymmetric matter, especially in the important cases $\cN=3$ and 
$\cN=4$.
In order to describe general supergravity-matter systems in these cases, superspace approaches 
appear to be the most useful ones. 

As is well known, a universal approach to engineering supergravity theories in diverse 
dimensions is 
to realise them as conformal supergravity coupled to certain compensating supermultiplet(s) 
\cite{KT}.
Making use of the conformal supergravity constraints 
on the superspace torsion proposed in \cite{HIPT}, in our recent work \cite{KLT-M-2011}
the superspace geometry of 3D ${\cal N}$-extended conformal supergravity was
developed\footnote{The special cases of $\cN=8$ and $\cN=16$ conformal supergravity theories 
were independently  worked out in 
\cite{Howe:2004ib,CGN} and \cite{GH} respectively. Recently, new results on $\cN=8$ conformal supergravity and its applications 
have  appeared \cite{GH2,GH3}.  }
and then applied (building on the structure of off-shell superconformal $\s$-models in three 
dimensions
\cite{KPT-MvU-2011})
to construct general off-shell supergravity-matter couplings
for the cases ${\cal N} \leq 4$.  In order to illustrate how the formalism of \cite{KLT-M-2011}
can be used to describe matter-coupled AdS supergravity theories, 
the cases $p+q=2$ were studied in detail in \cite{KT-M-2011}. 
In particular, Ref. \cite{KT-M-2011} provided two dual  off-shell formulations for (1,1) AdS 
supergravity 
and one off-shell formulation for (2,0) AdS supergravity. The most general $\s$-model couplings
to (1,1) and (2,0) AdS supergravity theories were  constructed in  \cite{KT-M-2011} from first 
principles.
These results generalise those obtained earlier \cite{IT,DKSS} 
within the Chern-Simons approach \cite{AT}.\footnote{Ref. \cite{IT} constructed
only those $\s$-model couplings to (2,0) AdS supergravity in which the scalar fields are neutral 
under 
the gauged U(1) $R$-symmetry group. Ref. \cite{DKSS} studied locally supersymmetric $\s$-
models on homogeneous
spaces of the form $ G/H \times \rm U(1)$ in which the scalar fields are charged under the gauged 
U(1) $R$-symmetry group. Such $\s$-model couplings  to (2,0) AdS supergravity
are special cases of those constructed 
in \cite{KT-M-2011}.}

The present paper is devoted to new applications of the formalism developed in  
\cite{KLT-M-2011}. 
First of all, here we introduce $(p,q)$ AdS superspaces and study their geometric properties. 
Secondly, we develop an off-shell formalism for constructing rigid  supersymmetric theories  
in AdS with $p+q = 3$, and specifically concentrate on describing 
general supersymmetric nonlinear $\s$-models.  

Within the framework of   \cite{KLT-M-2011}, $(p,q)$ AdS superspace 
\bea
{\rm AdS}_{(3|p,q)} = \frac{ {\rm OSp} (p|2; {\mathbb R} ) \times  {\rm OSp} (q|2; {\mathbb R} ) } 
{ {\rm SL}( 2, {\mathbb R}) \times {\rm SO}(p) \times {\rm SO}(q)}
\eea
originates as  a maximally symmetric supergeometry with covariantly constant torsion and 
curvature 
generated by a symmetric  torsion $S^{IJ}= S^{JI}$, with the structure-group indices $I, J$ 
taking values from 1 to $\cN$.  
It turns out that $S^{IJ}$ is nonsingular, and the parameters $p$ and $q= \cN-p$ 
determine its  signature. The ordinary AdS space 
\bea
{\rm AdS}_{3} = \frac{ \rm SL( 2, {\mathbb R})  \times  SL( 2, {\mathbb R})  } 
{\rm SL( 2, {\mathbb R}) }
\eea
is the bosonic body of ${\rm AdS}_{(3|p,q)} $. The curvature of ${\rm AdS}_{3} $ is proportional to 
$S^2 = S^{IJ} S_{IJ} /\cN$ (with the structure-group indices  being raised and lowered using  
$\d^{IJ}$
and $\d_{IJ}$).
The Killing vector fields of ${\rm AdS}_{(3|p,q)}$  can be shown to generate the isometry group
${\rm OSp} (p|2; {\mathbb R} ) \times  {\rm OSp} (q|2; {\mathbb R} )$. Among the superspaces 
 ${\rm AdS}_{(3|p,q)}$ with $p+q = \cN$ fixed, 
the largest isometry group corresponds to
${\rm AdS}_{(3|\cN,0)} \equiv {\rm AdS}^{3|2\cN}$. 

In fact, starting from the superspace geometry of $\cN$-extended conformal supergravity 
\cite{KLT-M-2011}
and restricting the torsion to be covariantly constant and Lorentz invariant, a general AdS 
superspace 
solution for $\cN\geq 4$ includes not only the torsion $S^{IJ}$ described above but also 
a completely antisymmetric torsion $X^{IJKL} = X^{[IJKL]} $. It turns out that the latter may be 
non-zero 
only if $S^{IJ} =S \d^{IJ}$,  which means $p=\cN$ and $q=0$. Such solutions define new AdS 
superspaces,
${\rm AdS}^{3|2\cN}_{S,X}$, for which the isometry group is, in general,  a subgroup 
 of $\rm OSp (\cN|2; {\mathbb R} ) \times  SL (2, {\mathbb R} )$.

Why bother to study supersymmetric nonlinear $\s$-models in  ${\rm AdS}_{(3|p,q)} $? 
Part of our motivation comes from four dimensions. 
Recently, it has been realised that rigid supersymmetric field theories in $\rm AdS_4$ have 
drastically different properties compared to their counterparts  in Minkowski space 
\cite{BK_AdS_supercurrent,Adams:2011vw,FS,BKsigma1,BKsigma2,KN,BKLT-M}.
Analogous results apply in five dimensions \cite{BaggerXiong,BaggerLi}. It is therefore natural to
study the specific features of rigid supersymmetric field theories in $\rm AdS_3$.\footnote{Locally supersymmetric 
nonlinear $\s$-models in three dimensions were constructed in the on-shell component  approach in \cite{dWNT,dWHS}.} 
And then we can immediately see that the 3D story is much richer than the 4D one, 
for in 3D there exist several versions of $\cN$-extended AdS superspace. 
These superspaces have different isometry 
groups, and therefore they should allow different matter couplings.

This paper is organised as follows. In section 2 we introduce the $(p,q)$ AdS superspaces
and study their geometrical properties. In section 3 we prove that 
all $(p,q)$ AdS superspaces with $X^{IJKL}=0$ are conformally flat. 
The specific features of the $(p,q)$ AdS superspaces 
with $p+q \leq 4$ are studied in section 4. 
In section 5 we develop a general setup to construct (3,0) and (2,1) supersymmetric theories
in AdS. Specifically, we define a family of covariant projective supermultiplets to describe 
supersymmetric
matter, and then present a manifestly supersymmetric action. We also give an expression for the 
action 
obtained by integrating out the superspace Grassmann variables.
In section 6 we demonstrate how to reformulate any  (3,0) and (2,1) supersymmetric field theory 
in AdS as a dynamical system in a certain $\cN=2$ AdS superspace. 
In section 7 we construct general off-shell supersymmetric $\s$-models in AdS. 
Section 8 contains a brief discussion of the  results obtained. 
Some details on the derivation of the component action (\ref{components-Ac}) are collected in the 
appendix. 

\section{Three-dimensional $(p,q)$ AdS superspaces}
\setcounter{equation}{0}

In this section, we develop the differential geometry of three-dimensional 
$\cN$-extended $(p,q)$ AdS superspaces.

\subsection{Superspace geometry of $\cN$-extended conformal supergravity}

All $(p,q)$ AdS superspaces can be realised as special background configurations 
within the 3D $\cN$-extended conformal supergravity that was originally sketched in \cite{HIPT} and 
then fully 
developed in \cite{KLT-M-2011}. In this subsection we recall those results of  \cite{KLT-M-2011}
which are necessary for our subsequent analysis.

In three dimensions, $\cN$-extended conformal supergravity can be described using a curved 
superspace which is parametrized by real   bosonic ($x^m$) 
and real fermionic ($\q^{\mu}_{\tt I}$) coordinates, 
\bea
z^{M}=(x^m,\q^{\mu}_{\tt I})~,
 \qquad m=0,1,2
~,~~~
\mu=1,2
~,~~~
{\tt I}
={\bf 1},\cdots,{\bm \cN}~,
\eea
and is characterised by the structure group
${\rm SL}(2,{\mathbb R}) \times {\rm SO} (\cN)$.
The  superspace differential geometry is encoded in
covariant derivatives of the form
\bea
\cD_{A}&\equiv& (\cD_a, \cD^I_\a )= 
E_{A}
+\O_A
+\F_A
~,
\label{cov-dev-000}
\eea
where the tangent space indices take the values
$\a=1,2$, $a=0,1,2$, $I={\bm 1},\cdots,{\bm \cN}$.
In eq. (\ref{cov-dev-000}),  
$E_A=E_A{}^M \pa_M$ is the supervielbein, with $\pa_M=\pa/\pa z^M$; 
\bea
\O_A=\hf\O_{A}{}^{bc}\cM_{bc}=-\O_{A}{}^b\cM_b=\hf\O_{A}{}^{\b\g}\cM_{\b\g}~,
~~~~
\cM_{ab}=-\cM_{ba}~,~~\cM_{\a\b}=\cM_{\b\a}
\eea
is the Lorentz connection;
and
\bea
\F_A=\hf\F_A{}^{KL}\cN_{KL}~,~~~~\cN_{KL}=-\cN_{LK}
\eea
is the SO($\cN$)-connection.
The Lorentz generators with two vector indices ($\cM_{ab}$), with one vector index ($\cM_a$)
and with two spinor indices ($\cM_{\a\b}$) are related to each other by the rules:
$\cM_a=\hf \ve_{abc}\cM^{bc}$ and $\cM_{\a\b}=(\g^a)_{\a\b}\cM_a$.
The generators of the group SO$(\cN)$ are denoted by $\cN_{IJ}$.
For more details on our notation and conventions see Appendix A of \cite{KLT-M-2011}.
The generators of SL(2,${\mathbb R}$)$\times$SO($\cN$) act on the covariant derivatives as 
follows:\footnote{The operation of (anti) symmetrization of $n$ indices is defined to involve 
a factor of $(n!)^{-1}$.}
\bsubeq
\bea
&&{\big [}{\cal M}_{ab} ,\cD_{\a}^{I}{\big ]} =\hf \ve_{abc}(\g^c)_\a{}^\b\cD_\b^{I}~,~~
{\big [}  {\cal M}_a,\cD_{\a}^{I} {\big]} =-\hf(\g_a)_\a{}^\b\cD_\b^{I} 
~,~~ 
~~~~~~~
\\
&&
{\big [}{\cal M}_{\a\b} ,\cD_{\g}^{I}{\big ]} =\ve_{\g(\a}\cD_{\b)}^{I}~,~~~
{\big [} {\cal M}_{ab} , \cD_{c}  {\big]} =2\eta_{c[a}\cD_{b]}~,~~~
{\big [} {\cal M}_{a} , \cD_{b}  {\big]} =\ve_{abc}\cD^{c}~,~~~~~~
\label{acM}
\\
&&
{\big [} {\cal N}{}_{KL},\cD_{\a}^{I}{\big]} =2\d^I_{[K}\cD_{\a L]} ~,~~
{\big [} {\cal N}{}_{KL},\cD_{a}{\big]} =0 ~.
~~~~~~~~~
\eea
\esubeq

To describe conformal supergravity,  
the covariant derivatives have to obey certain  constrains \cite{HIPT}.
With the constraints imposed, 
the Bianchi identities  lead to the following
(anti) commutation relations\footnote{For the purposes of this paper, we only need the explicit 
expressions
for the dimension-1 components of the  torsion and the curvature.} 
derived in  \cite{KLT-M-2011}:
\bsubeq\label{2.6}
\bea
\{\cD_\a^I,\cD_\b^J\}&=&
2\ri\d^{IJ}\cD_{\a\b}
-2\ri\ve_{\a\b}C^{\g\d}{}^{IJ}\cM_{\g\d}
-4\ri S^{IJ}\cM_{\a\b}
\non\\
&&
+\Big(
\ri\ve_{\a\b}X^{IJKL}
-4\ri\ve_{\a\b}S^{K}{}^{[I}\d^{J]L}
+\ri C_{\a\b}{}^{KL}\d^{IJ}
-4\ri C_{\a\b}{}^{K(I}\d^{J)L}
\Big)
\cN_{KL}
~,~~~~~~~~~
\label{alg-1}
\\
%%%%%%%%%%%%
{[}\cD_{\a\b},\cD_\g^K{]}
&=&
-\Big(
\ve_{\g(\a}C_{\b)\d}{}^{KL}
+\ve_{\d(\a}C_{\b)\g}{}^{KL}
+2\ve_{\g(\a}\ve_{\b)\d}S^{KL}
\Big)
\cD^\d_L
\non\\
&&
+\hf R_{\a\b}{}_\g^K{}^{de}\cM_{de}
+\hf R_{\a\b}{}_\g^K{}^{PQ}\cN_{PQ}
\label{alg-3/2}
~.
\eea
\esubeq
This algebra  is given in terms of  three
dimension-1 tensor superfields, $ X^{IJKL}$, $S^{IJ}$ and $C_a{}^{IJ}$,
which are real and have the symmetry properties
\bea
X^{IJKL}=X^{[IJKL]}~,~~~~
S^{IJ}=S^{(IJ)}
~,~~~~
C_a{}^{IJ}=C_a{}^{[IJ]}
~.
\label{dim-1-superfields}
\eea
The dimension-3/2 components of the curvature in (\ref{alg-3/2}),  
$R_{\a\b}{}_\g^K{}^{de}$ and $R_{\a\b}{}_\g^K{}^{PQ}$, 
are known algebraic functions  \cite{KLT-M-2011}
of  first spinor covariant derivatives of the 
dimension-1 tensor superfields
(\ref{dim-1-superfields}).
It is useful to represent $S^{IJ}$ as a sum of its irreducible components\footnote{In 
this paper, we make use of $ S^{IJ} $ and $ \cS^{IJ} $, as well as $\cS$ and 
$S:=\sqrt{  \d_{I K} \d_{J L} S^{ I J }  S^{ K L } / \cN}.$ We hope our imperfect notation 
will not lead to any confusion. }
\bea
S^{IJ}=\cS^{IJ}+\d^{IJ}\cS~,~~~~~~
\d_{KL}\cS^{KL}=0~,~~~
\cS:=\frac{1}{\cN}\d_{KL}S^{KL}
~.
\eea

The Bianchi identities imply the following set of differential equations   \cite{KLT-M-2011}
\bsubeq
\bea
\cD_\a^{I} \cS^{JK}&=&
2\cT_{\a}{}^{I(JK)}
+\cS_\a{}^{(J}\d^{K)I}
-{1\over \cN}\cS_\a{}^{I}\d^{JK}~,
\label{3/2-1} \\
%%%%%%%%%%%%
\cD_{\a}^{I} C_{\b\g}{}^{JK}
&=&
{2\over 3}\ve_{\a(\b}\Big(
C_{\g)}{}^{IJK}
+3\cT_{\g)}{}^{JKI}
+4(\cD_{\g)}^{[J} \cS)\d^{K]I}
+{(\cN-4)\over \cN}\cS_{\g)}{}^{[J}\d^{K]I}
\Big)
\non\\
&&
+C_{\a\b\g}{}^{IJK}
-2 C_{\a\b\g}{}^{[J}\d^{K]I}
~,
\label{3/2-2}\\
%%%%%%%%%%%%
\cD_\a^{I}X^{JKLP}
&=&
X_\a{}^{IJKLP}-4C_{\a}{}^{[JKL}\d^{P]I}
\label{3/2-3}
~,
\eea
\esubeq
where the dimension-3/2 superfields satisfy:
$\cT_{\a}{}^{IJK}=\cT_{\a}{}^{[IJ]K}$,
$\d_{JK}\cT_{\a}{}^{IJK}=\cT_{\a}{}^{[IJK]}=0$,
$C_{\a\b\g}{}^{IJK}=C_{(\a\b\g)}{}^{[IJK]}$,
$C_{\a\b\g}{}^{I}=C_{(\a\b\g)}{}^{I},$
$C_{\a}{}^{IJK}=C_{\a}{}^{[IJK]}$,
$X_\a{}^{IJKPQ}=X_\a{}^{[IJKPQ]}$.

 The supergravity gauge group is generated by local transformations of the form
\bea
\d_K\cD_A=[K,\cD_A]~,~~~~~~
K=K^C  \cD_C+\hf K^{cd} \cM_{cd}+\hf K^{PQ} \cN_{PQ}
~,
\label{SUGRA-gauge-group1}
\eea
with all the gauge parameters obeying natural reality conditions but otherwise arbitrary.
Given a tensor superfield $T$, it transforms as follows:
\bea
\d_KT=KT ~.
\label{SUGRA-gauge-group2}
\eea

The conformal supergravity constraints proposed in \cite{HIPT} are invariant under super-Weyl transformations. 
This invariance plays a key role in the discussion of the multiplet structure of $\cN$-extended 
conformal supergravity in 
\cite{HIPT}. The super-Weyl transformations 
were not given explicitly in \cite{HIPT}.  
For $\cN=8$ conformal supergravity, the finite form of super-Weyl transformations first appeared in \cite{Howe:2004ib}.
In the case of $\cN$-extended conformal supergravity, the infinitesimal form of these transformations was described in \cite{KLT-M-2011}.  Here we present for the first time the finite form of the $\cN$-extended super-Weyl transformations \cite{KLT-M-2011}.  
This result is essential for the analysis in section \ref{Conf-Flat}.

The super-Weyl transformation of the covariant derivatives is
\bsubeq
\bea
\cD'{}_\a^I&=&
\re^{\hf \s}\Big(\cD_\a^I + (\cD^{\b I}\s)\cM_{\a\b}+(\cD_{\a J} \s)\cN^{IJ}\Big)
~,
\label{Finite-sW-1-1}
\\
%%%
\cD'{}_{a}&=&
\re^{\s}\Big(\cD_{a}
+\frac{\ri}{2}(\g_a)^{\a\b}(\cD_{(\a}^{K} \s)\cD_{\b)K} 
+\ve_{abc}(\cD^b \s)\cM^c
+\frac{\ri}{16} (\g_a)^{\a\b}({[}\cD_{(\a}^{[K},\cD_{\b)}^{L]}{]}\s)\cN_{KL}
\non\\
&&~~~~~~
-\frac{\ri}{8}(\g_a)^{\a\b}(\cD^\r_{K} \s)(\cD_\r^{K}\s)\cM_{\a\b}
+\frac{3\ri}{8}(\g_a)^{\a\b}(\cD_{(\a}^{[K} \s)(\cD_{\b)}^{L]} \s)\cN_{KL}
\Big)
~,
\label{Finite-sW-1-2}
\eea
\esubeq
while the dimension-1 torsion and curvature tensors transform as
\bsubeq
\bea
S'{}^{IJ}&=&
\re^{\s}\Big(
S^{IJ}
- \frac{\ri}{8}({[}\cD^{\r (I},\cD_\r^{J)}{]}\s)
+\frac{\ri}{4}(\cD^{\r(I} \s)(\cD_\r^{J)}\s)
-\frac{\ri}{8}\d^{IJ}(\cD^\r_{K} \s)(\cD_\r^{K}\s)
\Big)
~,~~~~~~~~
\label{Finite-sW-2-1}
\\
%%%
C'{}_{a}{}^{IJ}&=&
\re^{\s}\Big(
C_{a}{}^{IJ}
-\frac{\ri}{8}(\g_a)^{\a\b}({[}\cD_{(\a}^{[I},\cD_{\b)}^{J]}{]}\s)
-\frac{\ri}{4}(\g_a)^{\a\b}(\cD_{(\a}^{[I} \s)(\cD_{\b)}^{J]}\s)
\Big)~,
\label{Finite-sW-2-2}
\\
%%%
X'{}^{IJKL}&=&
\re^{\s}X^{IJKL}
~.
\label{Finite-sW-2-3}
\eea
\esubeq
For later use, we rewrite the super-Weyl transformations of $S^{IJ}$ and $C_a{}^{IJ}$ 
in the following equivalent form:
\bsubeq
\bea
S'{}^{IJ}&=&
\Big(
\re^{\s}S^{IJ}
- \frac{\ri}{4}(\cD^{\r (I}\cD_\r^{J)}\re^{\s})
+\frac{\ri}{2}\re^{-\s}\big(\d^I_K\d^J_L-\frac{1}{4}\d^{IJ}\d_{KL}\big)
(\cD^{\r(K} \re^{\s})(\cD_\r^{L)}\re^{\s})
\Big)
~,~~~~~~~~~~~~
\label{Finite-sW-3-1}
\\
%%%
C'{}_{a}{}^{IJ}&=&
\Big(
C_{a}{}^{IJ}
-\frac{\ri}{4}(\g_a)^{\a\b}\cD_{(\a}^{[I}\cD_{\b)}^{J]}
\Big)\re^{\s}
~.
\label{Finite-sW-3-2}
\eea
\esubeq

This concludes 
our summary of the superspace geometry of $\cN$-extended conformal supergravity
\cite{KLT-M-2011}.

\subsection{Definition of $(p,q)$ AdS superspaces}

We are now prepared to introduce  AdS superspaces.
By definition, they correspond to those conformal supergravity backgrounds which 
satisfy the following requirements:

(i) the torsion and curvature tensors are Lorentz invariant;

(ii) the torsion and curvature tensors are covariantly constant.

\noindent
Condition (i) implies 
\bea
C_a{}^{IJ}\equiv 0
~.
\eea
Requirement (ii) has a series of implications.
First of all, the conditions
\bea
\cD_\a^I S^{JK}=\cD_a S^{JK}=0~,~~~
\cD_\a^I X^{JKLM}=\cD_a X^{JKLM}=0
\label{cov-const-S_X}
\eea
imply that all the dimension-3/2 curvatures in the second line of (\ref{alg-3/2}) are identically zero.
Moreover, the integrability conditions for the constraints (\ref{cov-const-S_X}), 
\bea
\{\cD_\a^I,\cD_\b^J\} S^{KL}=0~,~~~
\{\cD_\a^I,\cD_\b^J\} X^{KLMN}=0
~,
\eea
are equivalent to the following algebraic constraints on $S^{IJ}$ and $X^{IJKL}$:
\bsubeq
\bea
0&=&
X^{IJN(K}S_{N}{}^{L)}
-S^{IM}S_{M}{}^{(K}\d^{L)J}
+S^{JM}S_{M}{}^{(K}\d^{L)I}
\label{alg-constr-01}
~,
\\
0&=&
X^{IJN[K}X_{N}{}^{LPQ]}
+2S^{M}{}^{[I}\d^{J]N}\d^{[K}_{M}X_{N}{}^{LPQ]}
-2S^{M}{}^{[I}\d^{J]N}\d^{[K}_{N}X_{M}{}^{LPQ]}
~.
\label{alg-constr-02}
\eea
\esubeq

We now have to analyse all the implications of 
the Bianchi identities
\bea
\sum_{[ABC)}{[}\cD_A,{[}\cD_B,\cD_C\}\}=0
\label{Bianchi}
\eea
in the case that the covariant derivatives obeying 
the (anti) commutation relations (\ref{alg-1})--(\ref{alg-3/2})
are further subject  to the constraints (i) and (ii).
Solving the Bianchi identities
is straightforward albeit somewhat tedious and we omit the details.
By analysing (\ref{Bianchi}) we obtain a new crucial
constraint on the torsion tensor $S^{IJ}$:
\bea
{\hat S}^2 =S^2 {\mathbbm 1}~, \qquad
{\hat S}:= (S^{IJ} ) ={\hat S}^{\rm T} ~, \qquad
S^2 :=\frac{1}{\cN}\, \tr ({\hat S}^2) \geq0
~.
\label{alg-constr-SS}
\eea
This shows that $\hat S$ is a nonsingular symmetric $\cN \times \cN$ matrix if $S^2 >0$;
in this case ${\hat S}/S$ is an orthogonal matrix.
Moreover, by solving the Bianchi identities
one readily  derives a commutator of two vector covariant 
derivatives.
The complete algebra of covariant derivatives turns out to be
\bsubeq\label{2.16}
\bea
\{\cD_\a^I,\cD_\b^J\}&=&
2\ri\d^{IJ}\cD_{\a\b}
-4\ri S^{IJ}\cM_{\a\b}
+\ri\ve_{\a\b}\Big(
X^{IJKL}
-4S^{K}{}^{[I}\d^{J]L}
\Big)
\cN_{KL}
~,
\label{alg-AdS-1}
\\
%%%%%%%%%%%%
{[}\cD_{a},\cD_\b^J{]}
&=&
S^{J}{}_{K}(\g_a)_\b{}^\g\cD_{\g}^K
~,
\label{alg-AdS-3/2}
\\
%%%%%%%%%%%%
{[}\cD_{a},\cD_b{]}
&=&
4\,S^2\,
\ve_{abc}\cM^{c}
=
-4\,S^2\,\cM_{ab}
~.
\label{alg-AdS-2}
\eea
\esubeq
We recall that  the structure-group indices  are raised and lowered using  $\d^{IJ}$
and $\d_{IJ}$. 
Due to  (\ref{alg-constr-SS}), the constraints (\ref{alg-constr-01})--(\ref{alg-constr-02})
become
\bsubeq
\label{2.17}
\bea
0&=&
S^{(K}{}_{N}X^{L)IJN}~,
\label{alg-constr-2}
\\
0&=&
X_N{}^{IJ[K}X^{LPQ]N}
+S^{I[K}X^{LPQ]J}
-S^{J[K}X^{LPQ]I}
\non\\
&&
-S^{IM}X_{M}{}^{[LPQ}\d^{K]J}
+S^{JM}X_{M}{}^{[LPQ}\d^{K]I}
~.
\label{alg-constr-3}
\eea
\esubeq

In accordance with  (\ref{alg-constr-SS}), there are two conceptually different cases: 
(a) $S>0$; and (b) $S=0$ and hence  $S^{IJ}=0$. 
In the former case, eq. (\ref{alg-AdS-2}) tells us that 
the commutator of two vector covariant derivatives 
is exactly that of 3D AdS space. 
The latter case corresponds to a flat superspace with the following algebra of covariant 
derivatives:
\bsubeq\label{2.18}
\bea
\{\cD_\a^I,\cD_\b^J\}&=&
2\ri\d^{IJ}\cD_{\a\b}+\ri\ve_{\a\b}
X^{IJKL} \cN_{KL} ~,\\
%%%%%%%%%%%%
{[}\cD_{a},\cD_\b^J{]}
&=& 0~, 
\qquad
{[}\cD_{a},\cD_b{]} =0
~.
\eea
\esubeq
This superspace is of Minkowski type for $\cN=1,2,3$. However, for $\cN\geq 4$ 
there may exist a non-zero constant antisymmetric tensor $X^{IJKL}$ constrained by 
\bea
X_N{}^{IJ[K}X^{LPQ]N} =0 ~, 
\label{2.19}
\eea
so that the resulting superspace is a deformation of $\cN$--extended Minkowski superspace. 
In what follows, our analysis will be restricted to the AdS case, $S>0$.

%%%%%%%%%%%%%%%%%%%%%%%%%%%%%%%%%%%%%%%%%%%%%%%

\subsection{Analysis of the $(p,q)$ AdS constraints}

We have seen that the 
(anti) commutation relations \eqref{2.16}
 require the algebraic constraints (\ref{alg-constr-SS}) and \eqref{2.17}
 as the consistency conditions.
Let us analyse the implications of these equations.
The most important equation to study is (\ref{alg-constr-SS}).

The  torsion ${\hat S}=(S^{IJ})$ is a real symmetric $\cN\times\cN$ matrix. 
A local  SO($\cN$) transformation can be performed  to diagonalise $\hat S$.
Then, without loss of generality, the general solution of the constraint (\ref{alg-constr-SS}) is
\bea
S^{IJ}=S\,{\rm{diag}}(\,  \overbrace{+1,\cdots,+1}^{p} \, , \overbrace{-1,\cdots,-1}^{q=\cN-p} \,)
~,
\label{diag-S}
\eea
where $S=\sqrt{(S^{IJ}S_{IJ})/\cN} >0$ is a positive parameter of unit dimension. 
In the `diagonal frame' (\ref{diag-S}), we are left with an unbroken local group 
${\rm SO}(p)\times {\rm SO}(q)$.  
In what follows, we assume $p\geq q$.
Such a superspace should originate as a maximally symmetric solution of the $(p,q)$ AdS 
supergravity. 
The integers $p$ and $q$ determine the signature of $S^{IJ}$.

For our subsequent analysis, it is handy to introduce a special notation associated
with the diagonal frame (\ref{diag-S}). 
All the isovector indices running from $1$ to $p$ will be overlined, 
while those taking values from $p+1$ to $\cN$ will be underlined. 
With this notation, 
 the components of  $S^{IJ}$ in the diagonal frame are
 \bea
S^{\overline{I}\overline{J}}=S\,\d^{\overline{I}\overline{J}}~,~~~
S^{\underline{I}\underline{J}}=-S\,\d^{\underline{I}\underline{J}}~,~~~
S^{\OI\UJ}=S^{\UI\OJ}=0~.
\label{diag-frame}
\eea
The diagonal frame is especially useful for solving  the constraints obeyed by  $X^{IJKL}$.

Making use of  (\ref{diag-frame}), 
it is easy to see that the constraint 
(\ref{alg-constr-2}) is equivalent to
\bea
X^{IJ\UK\OL}=0
~.
\label{cond-X-1}
\eea
This means that the only non-zero components of $X^{IJKL}$ 
are those which have all the indices of the same type, i.e.
$X^{\overline{IJKL}}$ and $X^{\underline{IJKL}}$.

Using (\ref{diag-frame}) and (\ref{cond-X-1}),  
the second constraint on $X^{IJKL}$, eq. 
(\ref{alg-constr-3}), dramatically  simplifies.
The strongest condition arises when one chooses
the index  $I$ overlined and the index $J$ underlined. In this case, due to
 (\ref{diag-frame}) and (\ref{cond-X-1}), the expression (\ref{alg-constr-3}) is non-trivial only if, 
 among the indices $K,\,L,\,P$ and $Q$, one is overlined and the other three are underlined or
vice versa. 
We then get the following equations
\bsubeq
\bea
0&=&
S\,X^{\UJ\UL\UP\UQ}\d^{\OK\OI}
~,
\\
0&=&
S\,X^{\OI\OL\OP\OQ}\d^{\UK\UJ}
~.
\eea
\esubeq
It is clear that these equations can have nontrivial solutions  only in the $(\cN,0)$ case.
We have thus proved that the curvature $X^{IJKL}$  can only consistently
appear in the AdS algebra  if $S^{IJ}=S\,\d^{IJ}$.
In this case, the equation (\ref{alg-constr-3}) simplifies to
\bea
X_N{}^{IJ[K}X^{LPQ]N}=0
~.
\label{fin-constr-X}
\eea
This is the same algebraic equation which  emerges in the case of deformed Minkowski 
superspace, eq. \eqref{2.19}. 

The first case where $X^{IJKL}$ can appear in the algebra is $\cN=4$. Here 
\bea
X^{IJKL}=X\ve^{IJKL}
~,
\eea
where $X$ is a real constant parameter, and
 $\ve^{IJKL}$  
the completely antisymmetric Levi-Civita tensor
(normalised by 
$\ve$${}^{{\footnotesize{\bm 1}{\footnotesize{\bm 2}}{\footnotesize{\bm 3}}{\footnotesize{\bm 4}}}}$$\,=1$) 
which is invariant under SO(4).
The constraint (\ref{fin-constr-X}) is automatically satisfied due to the identity
\bea
\ve_{IJKP}\ve^{LMNP}&=&6\,\d_{[I}^L\d_J^M\d_{K]}^N
~.
\eea

Here we do not give a general solution of eq. \eqref{fin-constr-X} for $\cN>4$.
We just mention that a particular   solution of eq.  \eqref{fin-constr-X}  for any $\cN>4$ is 
obtained by choosing
$X^{IJKL}=X\ve^{IJKL}$${}^{\,{\footnotesize{\bm 5}}\,{\footnotesize{\bm 6}}
\cdots\,{\footnotesize{\bm \cN}}}$, with $\ve^{I_1 \dots I_{\cN} }$ the appropriate Levi-Civita tensor. 
This  solution is invariant under 
a subgroup ${\rm SO}(4) \times {\rm SO}(\cN-4)$ of the gauged $R$-symmetry   group SO$(\cN)$.

We conclude by rewriting the algebra of covariant derivatives 
(\ref{alg-AdS-1})--(\ref{alg-AdS-2}) 
in the diagonal frame (\ref{diag-S})
for  general $(p,q)$  
with $q>0$ (in the  case $(\cN,0)$
the algebra of covariant derivatives is given by eqs. 
(\ref{alg-AdS-1})--(\ref{alg-AdS-2}) with $S^{IJ}=S\,\d^{IJ}$)
\bsubeq
\bea
\{\cD_\a^\OI,\cD_\b^\OJ\}&=&
2\ri\d^{\OI\OJ}\cD_{\a\b}
-4\ri \,S\,\d^{\OI\OJ}\cM_{\a\b}
-4\ri\,S\,\ve_{\a\b}\cN^{\OI\OJ}
~,
\label{alg-AdS-2-1a}
\\
%%%%%%%%%%%%
\{\cD_\a^\UI,\cD_\b^\UJ\}&=&
2\ri\d^{\UI\UJ}\cD_{\a\b}
+4\ri \,S \,\d^{\UI\UJ}\cM_{\a\b}
+4\ri\,S\,\ve_{\a\b}\cN^{\UI\UJ}
~,
\label{alg-AdS-2-1b}
\\
%%%%%%%%%%%%
\{\cD_\a^\OI,\cD_\b^\UJ\}&=&
\{\cD_\a^\UI,\cD_\b^\OJ\}=
0
~,
\label{alg-AdS-2-1}
\\
%%%%%%%%%%%%
{[}\cD_{a},\cD_\b^\OJ{]}
&=&
S\,(\g_a)_\b{}^\g\cD_{\g}^\OJ
~,
~~~~~~
{[}\cD_{a},\cD_\b^\UJ{]}
=
-S\,(\g_a)_\b{}^\g\cD_{\g}^\UJ
~,
\label{alg-AdS-2-3/2}
\\
%%%%%%%%%%%%
{[}\cD_{a},\cD_b{]}
&=&
4\,S^2\,
\ve_{abc}\cM^{c}
=
-\,4\,S^2\,\cM_{ab}
~.
\label{alg-AdS-2-2}
\eea
\esubeq
Note that the $R$-symmetry group of this superspace is 
SO($p$)$\,\times\,$SO($q$).

\subsection{The  Killing vector fields of $(p,q)$ AdS superspace}
\label{Killings-subsection}

To describe rigid supersymmetric field theories in $(p,q)$ AdS superspaces,
we need to develop a superfield description of the corresponding isometry transformations. 
Here we use the diagonal frame where the results become more transparent, and consider only 
the cases $X^{IJKL}=0$.
The isometry transformations are generated by $(p,q)$ AdS Killing vector fields, 
\bea 
\x=\x^a\cD_a
+\x^{\a}_\OI\cD_\a^\OI
+\x^{\a}_\UI\cD_\a^\UI
~,
\label{2.28}
\eea
which by definition obey the equation
\bea
\Big{[}\x
+\hf\L^{\OI\OJ}\cN_{\OI\OJ}
+\hf\L^{\UI\UJ}\cN_{\UI\UJ}
+\hf\L^{ab}\cM_{ab}
,\cD_C\Big{]}=0~,
\label{2.29}
\eea
for some parameters $\L^{\OI\OJ}$, $\L^{\UI\UJ}$ and $\L^{ab}$.
This equation is equivalent to the relations 
\bsubeq\label{2.30}
\bea
&\cD_\a^{\OI}\x_\b^{\OJ}
=
-\ve_{\a\b}\L^{\OI\OJ}+S\d^{\OI\OJ}\x_{\a\b}+\hf\d^{\OI\OJ}\L_{\a\b}
~,
~~~~~~
\cD_\a^\UI\x_\b^\OJ=0
~,
\\
%%%%%%
&\cD_\a^{\UI}\x_\b^{\UJ}
=
-\ve_{\a\b}\L^{\UI\UJ}-S\d^{\UI\UJ}\x_{\a\b}+\hf\d^{\UI\UJ}\L_{\a\b}
~,~~~~~~
\cD_\a^\OI\x_\b^\UJ=0
~,
\\
%%%%%%
&0=
\cD_{ \g \OI}\x^{\a\g}
+6\ri\x^{\a}_{\OI}
~,~~~~~~
0=
\cD_{ \g \UI}\x^{\a\g}
+6\ri\x^{\a}_{\UI}
~,
\\
%%%%%%
&0=
\cD_\g^{\OI}\L^{\a\g}
+12\ri\,S\,\x^{\a\OI}
~,~~~~~~
0=
\cD_\g^{\UI}\L^{\a\g}
-12\ri\,S\,\x^{\a \UI}
~,
%%%%%%
\\
&0=
\cD_{(\a}^\OI\x_{\b\g)}
=\cD_{(\a}^\UI\x_{\b\g)}
=
\cD_{(\a}^{\OI}\L_{\b\g)}
=\cD_{(\a}^{\UI}\L_{\b\g)}
\label{K-07-2}
\eea
\esubeq
which imply the standard Killing vector equation
\bea
\cD_a\x_b+\cD_b\x_a=0~.
\eea
In accordance with \eqref{2.30}, the parameters $\x^{\a}_\OI $, 
$\x^{\a}_\UI $,  $\L^{\OI\OJ}$, $\L^{\UI\UJ}$ and $\L^{ab}$ are uniquely determined in terms of 
$\x^a$.
It can be shown that the $(p,q)$ AdS Killing vector fields 
generate the AdS supergroup 
${\rm OSp(}p|2;{\mathbb R}) \times {\rm OSp}(q|2;{\mathbb R})$.

%%%%%%%%%%%%%%%%%%%%%%%%%%%%%%%%%%%%%%%%%%%%%%%%
%%%%%%%%%%%%%%%%%%%%%%%%%%%%%%%%%%%%%%%%%%%%%%%%
%%%%%%%%%%%%%%%%%%%%%%%%%%%%%%%%%%%%%%%%%%%%%%%%

\section{Conformal flatness of $(p,q)$ AdS superspaces}
\setcounter{equation}{0}
\label{Conf-Flat}

It was demonstrated in \cite{KT-M-2011}  that AdS$_{(3|2,0)}$ and AdS$_{(3|1,1)}$ 
are conformally flat superspaces.
Here we generalise this result to the case of arbitrary $(p,q)$ AdS superspaces  with 
$X^{IJKL}=0$.
All superspaces AdS$_{(3|p,q)}$ are demonstrated to be  conformally flat.
Since  the super-Weyl transformation of $X^{IJKL}$ is homogeneous,  
any AdS superspace with  $X^{IJKL}\ne0$ is not conformally flat.

The super-Weyl transformations in $\cN$-extended conformal supergravity are given by eqs.
(\ref{Finite-sW-1-1})--(\ref{Finite-sW-1-2}).
Our goal is to show that there exists a local parametrisation of the superspace AdS$_{(3|p,q)}$ 
such that the covariant derivatives  $\cD_A$ take the form
\bsubeq \label{conf-flat}
\bea
\cD_\a^I&=&
\re^{\hf \s}\Big(D_\a^I + (D^{\b I}\s)\cM_{\a\b}+(D_{\a J} \s)\cN^{IJ}\Big)
~,
\label{conf-flat-1-1}
\\
%%%
\cD_{a}&=&
\re^{\s}\Big(\pa_{a}
+\frac{\ri}{2}(\g_a)^{\a\b}(D_{(\a}^{K} \s)D_{\b)K} 
+\ve_{abc}(\pa^b \s)\cM^c
+\frac{\ri}{16} (\g_a)^{\a\b}({[}D_{(\a}^{[K},D_{\b)}^{L]}{]}\s)\cN_{KL}
\non\\
&&~~~~~~
-\frac{\ri}{8}(\g_a)^{\a\b}(D^\r_{K} \s)(D_\r^{K}\s)\cM_{\a\b}
+\frac{3\ri}{8}(\g_a)^{\a\b}(D_{(\a}^{[K} \s)(D_{\b)}^{L]} \s)\cN_{KL}
\Big)
~,
\label{conf-flat-1-2}
\eea
\esubeq
for some real scalar $\s$.
 Here $D_A=(\pa_a,D_\a^I)$ are the covariant derivatives
of $\cN$-extended 3D Minkowski superspace, 
\bsubeq
\bea
\pa_a&=&\frac{\pa}{\pa x^a}~,~~~~~~
D_\a^I=\frac{\pa}{\pa\q^\a_I}+\ri(\g^a)_{\a\b}\q^{\b I}\pa_a~,
\eea
obeying the (anti) commutation relations
\bea
%%%%%%
\{D_\a^I, D_\b^J\}&=&2\ri\d^{IJ}(\g^a)_{\a\b}\pa_a~,~~~~~~
{[}\pa_a,D_\b^J{]}={[}\pa_a,\pa_b{]}=0
~.
\eea
\esubeq
Under the super-Weyl transformations, the dimension-1 torsion and curvature superfields
transform according to  the equations (\ref{Finite-sW-2-1})--(\ref{Finite-sW-2-3}).
The superspace AdS$_{(3|p,q)}$ is characterised by the conditions
$C_a{}^{KL}=X^{IJKL}=0$. 
Hence the parameter $\re^{\s}$ in \eqref{conf-flat}
must satisfy the equations 
\bsubeq
\bea
S^{IJ}&=&
- \frac{\ri}{4}(D^{\r (I}D_\r^{J)}\re^\s)
+ \frac{\ri}{2}\re^{-\s}(D^{\r (I}\re^\s)(D_\r^{J)}\re^\s)
-\frac{\ri}{8}\d^{IJ}\re^{-\s}(D^\r_{K} \re^\s)(D_\r^{K}\re^\s)
~,
\label{S-conf-flat}
\\
%%%%%%
0&=&D_{(\a}^{[I}D_{\b)}^{J]}\re^{\s}
~.
\label{C-0-conf-flat}
\eea
\esubeq
Moreover, in accordance with the analysis of the previous section, 
 the superfield $S^{IJ}$  has  to be covariantly constant, 
 \bea
 \cD_A S^{JK}=0~,
\label{3.444}
 \eea
and obey the algebraic constraint  (\ref{alg-constr-SS}) which we rewrite  as
\bea
S^{IK}S_{KJ}=S^2\d^{I}_{J}~,~~~~~~S^2=\frac{1}{\cN}S^{KL}S_{KL}
~.
\label{alg-constr-SS-2}
\eea
The equations  \eqref{3.444} and (\ref{alg-constr-SS-2}) 
have to be obeyed by  $\re^{\s}$ in addition to the condition \eqref{C-0-conf-flat}. 

To find a solution of the above equations, we make 
a  Lorentz invariant ansatz for the super-Weyl parameter
\bea
\re^{\s}&=&
1
+as^2x^2- \Q_s
+bs^2\Q^2+c\Q_s^2+ds\Q \Q_s
~,
\label{Ansatz-rewritten}
\eea
where
\bsubeq
\bea
x^2:=x^ax_a~,~~~~~~
\q^{IJ}&:=&\q^{\g I}\q_\g^{J}=\q^{JI}~,~~~
\Q:=\ri\d_{KL}\q^{KL}~,~~~
\Q_s:=\ri s_{KL}\q^{KL}
~,~~~~~~
\\
&&s_{IJ}=s_{JI}~,~~~~~~
s:=\sqrt{\frac{s^{KL}s_{KL}}{\cN}}
~.
\eea
\esubeq
The constant parameters
 $a,b,c$ and $d$ in (\ref{Ansatz-rewritten})
are real and dimensionless. As to 
the constant tensor $s_{IJ}$, it  is also  real 
and has unit mass dimension.  

The ansatz \eqref{Ansatz-rewritten} has been shown to be the correct one in the case of 
the (2,0) and (1,1) AdS superspace \cite{KT-M-2011}.  
It is also reminiscent of the conformally flat parametrisation of the 
4D $\cN$-extended AdS superspace 
derived in \cite{BILS} using group-theoretic techniques 
(direct proofs based on the use of super-Weyl transformations in 4D $\cN=1$  
and $\cN=2$ supergravity theories can be found, e.g., in 
 \cite{BK}  and \cite{KT-M_4D-conf-flat} respectively).

Let us turn to solving the conditions for conformal flatness using the ansatz introduced. 
As a first step, we observe from eq. (\ref{S-conf-flat}) that
\bea
S^{IJ}=s^{IJ}+\cO(\q)
~.
\eea
By considering the $\q$-independent part of (\ref{alg-constr-SS-2}),
we see  that $s^{IJ}$ has to be constrained by
\bea
s^{IK}s_{KJ}=\d^I_Ks^2
~.
\label{s-constr-1}
\eea
Next, we impose the equation (\ref{C-0-conf-flat}).
After some algebra, one sees that (\ref{C-0-conf-flat})  is satisfied provided 
\bea
b=-\frac{a}{4}~,~~~~~~
c=d=0
~.
\eea

To fix the value of $a$, it suffices to use again the equation (\ref{alg-constr-SS-2})
which  so far  has been solved at the $\q=0$ order only.
This equation tells us that 
\bea
a=-1
~.
\eea
We end up with the following expression for the super-Weyl parameter: 
\bea
\re^{\s}=1-s^2x^2-\Q_s+\frac{1}{4}s^2\Q^2
=1-s^2x^2-\ri s^{KL}\q_{KL}-\frac{1}{8}s^2(\d^{KL}\q_{KL})^2
~.~~~~~~~~~
\label{sigma-solution}
\eea
The geometry is characterized by the torsion $S^{IJ}$ that,
by using (\ref{S-conf-flat}), can be computed to be
\bea
S^{IJ}=
s^{IJ}
 + 2\ri 
 \frac{
s^2\q^{IJ}-s^{K(I}s^{J)L}\q_{KL}
+2 s^2s^{K(I}\q_\g^{ J)}\q_{\d K}x^{\g\d}
-s^2\q^{IJ}\Q_s
+s^2s^{K(I}\q^{J)}{}_K\Q
}
{1-s^2x^2-\Q_s+\frac{1}{4}s^2\Q^2}
~.~~~~~~
\label{S-solution}
\eea
It is an instructive exercise to 
prove the important relations
\bsubeq
\bea
\cS:=\frac{1}{\cN}\d_{KL}S^{KL}&=&\frac{1}{\cN}\d_{KL}s^{KL}~~~~~~\Longrightarrow~~~~~~
\cD_A\cS=0
~,
\label{3.144a}
\\
S^2&=&s^2~~~~~~\Longrightarrow~~~~~~
\cD_AS^2=0
~.
\label{3.144b}
\eea
\esubeq

To complete the analysis, we need to prove that 
the condition  \eqref{3.444} holds,
with the covariant derivatives defined by eqs. (\ref{conf-flat}).
This is equivalent to proving that $\cD_\a^{I}S^{JK}=0$.
We can simplify such a check by making a series of simple considerations.
First of all, due to  the very nature of the super-Weyl transformations,
the covariant derivatives \eqref{conf-flat} define 
some conformal supergravity background with the additional conditions:
the corresponding torsion and curvature tensors  satisfy 
the equations (\ref{3/2-1})--(\ref{3/2-3})
with $X^{IJKL}=C_a{}^{KL}=0$.
This means that  the  torsion $S^{IJ}=\cS^{IJ}+\d^{IJ}\cS$ in eq. (\ref{S-solution}) satisfies
\bsubeq
\bea
&&\cD_\a^{I} \cS^{JK}=
\cS_\a{}^{(J}\d^{K)I}
-{1\over \cN}\cS_\a{}^{I}\d^{JK}
\label{DS111}
~,
\\
%%%%%%%%%%%%
&&~~~
0
=
4\cN(\cD_{\a}^{I} \cS)
+(\cN-4)\cS_{\a}{}^{I}
~.
\label{3.155b}
\eea
\esubeq
It follows   that for $\cN\ne4$,  
a sufficient condition to have $\cD_\a^IS^{JK}=0$ 
is $\cD_\a^I\cS=0$. This is indeed the case in accordance with \eqref{3.144a}. 

In the $\cN=4$ case, the condition  $\cD_\a^I\cS=0$
follows from  \eqref{3.144a} and \eqref{3.155b}. 
However we need to independently check whether 
the requirement  $\cS_\a{}^I=0$ holds indeed. 
Using
$\cD_\a^I\cS=0$ and the representation $S^2=(\cS^{KL}\cS_{KL})/\cN+\cS^2$ gives
\bea
\cN\cD_\a^IS^2=2\big(
S^I{}_{J}\cS_\a{}^{J}
-\cS \cS_\a{}^{I}
\big)~.
\eea
Here the left-hand side is zero due to \eqref{3.144b}, and hence 
\bea
\cS\cS_\a{}^{I}=S^I{}_{J}\cS_\a{}^{J}
~.
\label{finalizing-0}
\eea
Since  $S^I{}_J$ is invertible, 
 this  equation gives $\cS_\a{}^I=0$ in the case that  $\cS=0$.
On the other hand, choosing  $\cS\ne0$ in  (\ref{finalizing-0}) gives
\bea
\cS_\a{}^{I}
=\Big(1+\frac{\cS^{KL}\cS_{KL}}{\cN\cS}\Big)\cS_\a{}^{I}
~.
\label{finalizing}
\eea
Ultimately this equation tells us  that $\cS_\a{}^I=0$.
Therefore, we have demonstrated  that $S^{IJ}$ is covariantly constant. 

We conclude with a comment about the space-time geometry 
associated  with the superspace AdS$_{(3|p,q)}$.
Given the expression for $\re^{\s}$, eq. \eqref{sigma-solution}, 
and the explicit form of the vector covariant derivative $\cD_a$, 
eq. (\ref{conf-flat-1-2}), we can read off  the space-time metric 
\be
{\rm d}s^2 = {\rm d}x^a  \,{\rm d}x_a \,\big(\re^{-2\s}\big) \big|_{\q=0}
= \frac{{\rm d}x^a  {\rm d}x_a }{\big(1-s^2 x^2\big)^2}~.
\label{metric}
\ee  
This coincides  with a standard expression for the metric  of AdS${}_3$
computed using the  stereographic projection for an AdS hyperboloid.\footnote{See, e.g,
Appendix D of \cite{KT-M_4D-conf-flat}
for details about the stereographic projection for ${\rm AdS}_d$.}

%%%%%%%%%%%%%%%%%%%%%%%%%%%%%%%%%%%%%%%%%%%%%%%%
%%%%%%%%%%%%%%%%%%%%%%%%%%%%%%%%%%%%%%%%%%%%%%%%
%%%%%%%%%%%%%%%%%%%%%%%%%%%%%%%%%%%%%%%%%%%%%%%%

\section{Elaborating on the AdS superspaces with $p+q \leq 4$}
\setcounter{equation}{0}

In this section we would like to reformulate the algebra of covariant derivatives, 
which corresponds to 
a given $(p,q)$ AdS superspace with $p+q\leq 4$, in a form that is more suitable 
for describing matter couplings within the supergravity formulation of \cite{KLT-M-2011}.

\subsection{$\cN=1$}

In the $\cN=1$ case,  only the $(1,0)$ AdS superspace is available. 
Its  geometry  is determined by the (anti) commutation relations
\bsubeq
\bea
\{\cD_\a,\cD_\b\}&=&
2\ri\cD_{\a\b}
-4\ri \,S\,\cM_{\a\b}
~,~~~~~~~~~
\label{N=1alg-1}
\\
%%%%%%%%%%%%
{[}\cD_{a},\cD_\b{]}
&=&
\,S\,(\g_a)_{\b}{}^{\g}\cD_{\g}
~,
\label{N=1alg-3/2}
%%%%%%%%%%%%
\\
{[}\cD_{a},\cD_b{]}
&=&
4\,\ve_{abc}\,S^2\cM^c
=-4\,S^2\cM_{ab}
~.
~~~~~~~~~~~~
\label{N=1alg-2}
\eea
\esubeq

%%%%%%%%%%%%%%%%%%%%%%%%%%%%%%%%%%%%%%%%%%%%%%%%

\subsection{$\cN=2$}

In the $\cN=2$ case,  there are two AdS superspaces: (2,0) and (1,1). 
They have already been studied in \cite{KT-M-2011}.
Here we would like to re-derive the main results of \cite{KT-M-2011}
using the analysis of the previous section.

\subsubsection{(2,0) AdS superspace}

The (2,0) covariant derivatives  satisfy the (anti) commutation relations
\bsubeq
\bea
&\{\cD_\a^I,\cD_\b^J\}=
2\ri\d^{IJ}\cD_{\a\b}
-4\ri \,S\,\d^{IJ}\cM_{\a\b}
+4\,\ve_{\a\b}\,S\,\ve^{IJ}\cJ
~,
\label{2_0-alg-AdS-1}
\\
%%%%%%%%%%%%
&{[}\cD_{a},\cD_\b^J{]}
=
\,S\,(\g_a)_\b{}^\g\cD_{\g}^J
~,~~~~~~
{[}\cD_{a},\cD_b{]}
=
-4\,S^2\,\cM_{ab}
~,
\label{2_0-alg-AdS-2}
\eea
\esubeq
where 
$\ve^{{\footnotesize{\bf 1}}\footnotesize{\bf 2}}=\ve_{{\footnotesize{\bf 1}}\footnotesize{\bf 2}}=1$ 
and we have introduced the U(1) generator $\cJ$ following  \cite{KLT-M-2011} 
\bea
\cN_{KL}=\ri\ve_{KL}\cJ~,~~~~
\cJ=-\frac{\ri}{2}\ve^{PQ}\cN_{PQ}~,
~~~~~~
{[}\cJ,\cD_\a^I{]}=-\ri\ve^{IJ}\cD_{\a J}~.
\eea
It is useful to switch to a complex basis for the spinor covariant derivatives,  
 $\cD_\a^I\to(\bfD_\a,\bfDB_\a)$,   such that 
 $\bfD_\a$ and $\bfDB_\a$ possess definite U(1) charges
 \bsubeq
 \bea
& \bfD_\a=\frac{1}{\sqrt{2}}(\cD_\a^{\bf 1}-\ri\cD_\a^{\bf 2})~,~~~
  \bfDB_\a=-\frac{1}{\sqrt{2}}(\cD_\a^{\bf 1}+\ri\cD_\a^{\bf 2})
  ~,
  \\
&  {[}\cJ,\bfD_\a{]}=\bfD_\a~,~~~
    {[}\cJ,\bfDB_\a{]}=-\bfDB_\a
    ~.
 \eea
\esubeq
With the definition $\bfD_a=\cD_a$, 
the algebra of covariant derivatives becomes
\bsubeq \label{20AdSsuperspace}
\bea
&\{\bfD_\a,\bfD_\b\}
=0
~,~~~~~~
\{\bfD_\a,\bfDB_\b\}
=
-2\ri\bfD_{\a\b}
-4\ri \,S\, \ve_{\a\b} \cJ
+4\ri \,S\, \cM_{\a\b} ~, 
\label{AdS_(2,0)_algebra_1}
\\
%%%%%%
&{[}\bfD_{a},\bfD_\b{]}
=
\,S\, (\g_a)_\b{}^\g\bfD_{\g}
~,~~~~~~
%%%%%%
{[}\bfD_a,\bfD_b{]} = -4\, S^2\, \cM_{ab}~,
~~~~~~
\label{AdS_(2,0)_algebra_2}
\eea
\esubeq
together with their complex conjugates.
Upon a redefinition of the AdS parameter,  $S\to \r /4$, 
these (anti) commutation relations become identical to those 
which define the (2,0) AdS 
geometry introduced in \cite{KT-M-2011}.

\subsubsection{(1,1) AdS superspace}

Now, let us turn to the (1,1) case. 
The algebra (\ref{alg-AdS-2-1a})--(\ref{alg-AdS-2-2}) becomes
\bsubeq
\bea
&\{\cD_\a^{\bf1},\cD_\b^{\bf1}\}=
2\ri\cD_{\a\b}
-4\ri \,S\,\cM_{\a\b}
~,
~~
\{\cD_\a^{\bf2},\cD_\b^{\bf2}\}=
2\ri\cD_{\a\b}
+4\ri \,S\, \cM_{\a\b}
~,~~
\{\cD_\a^{\bf1},\cD_\b^{\bf2}\}=0
~,~~~~~~~
\label{1_1-alg-AdS-2-1b}
\\
%%%%%%%%%%%%
&{[}\cD_{a},\cD_\b^{\bf1}{]}
=
\,S\,(\g_a)_\b{}^\g\cD_{\g}^{\bf1}
~,
~~
{[}\cD_{a},\cD_\b^{\bf2}{]}
=
-\,S\,(\g_a)_\b{}^\g\cD_{\g}^{\bf2}
~,~~~~
{[}\cD_{a},\cD_b{]}
=-4\,S^2\,\cM_{ab}
~.~~~~
\label{1_1-alg-AdS-2}
\eea
\esubeq
We can introduce a complex basis for the covariant derivatives defined by
 \bea
 \de_\a=\frac{\re^{\ri\vf}}{\sqrt{2}}(\cD_\a^{\bf 1}-\ri\cD_\a^{\bf 2})~,~~~
  \deb_\a=-\frac{\re^{-\ri\vf}}{\sqrt{2}}(\cD_\a^{\bf 1}+\ri\cD_\a^{\bf 2})~,
 \eea
with $\vf$ an arbitrary constant real phase.
Then, the  (anti) commutation relations  (\ref{1_1-alg-AdS-2-1b})--(\ref{1_1-alg-AdS-2}) turn into
\bsubeq \label{11AdSsuperspace}
\bea
&\{\de_\a,\de_\b\}
=
-4\bar{\mu}\cM_{\a\b}
~,~~~
 \{\deb_\a,\deb_\b\}
=
4\mu\cM_{\a\b}
~,~~~
\{\de_\a,\deb_\b\}
=
-2\ri\de_{\a\b}~, 
\label{AdS_(1,1)_algebra_1}
\\
%%%%%%
&{[}\de_{a},\de_\b{]}
=
\ri\bar{\mu}(\g_a)_\b{}^\g\deb_{\g}
~,~~
{[}\de_{a},\deb_\b{]}
=
-\ri\mu(\g_a)_\b{}^\g\de_{\g}
~,~~~~
{[}\de_a,\de_b]{}
= -4 |\mu|^2 \cM_{ab} ~,~~~
\label{AdS_(1,1)_algebra_2}
\eea
\esubeq
where
\bea
\mu:=-\,\ri\,\re^{2\ri\vf}\, S~.
\eea
This is exactly the  algebra of (1,1) AdS covariant derivatives \cite{KT-M-2011} .

%%%%%%%%%%%%%%%%%%%%%%%%%%%%%%%%%%%%%%%%%%%%%%%

\subsection{$\cN=3$}

There are two AdS superspaces in the  $\cN=3$ case: (3,0) and (2,1).

\subsubsection{(3,0) AdS superspace}
Let us start with the (3,0) AdS geometry described by 
\bsubeq
\bea
&\{\cD_\a^I,\cD_\b^J\}=
2\ri\d^{IJ}\cD_{\a\b}
-4\ri \,S\,\d^{IJ}\cM_{\a\b}
-4\ri\,S\,\ve_{\a\b}\cN^{IJ}
~,
\label{3_0-alg-AdS-1}
\\
%%%%%%%%%%%%
&{[}\cD_{a},\cD_\b^J{]}
=
\,S\,(\g_a)_\b{}^\g\cD_{\g}^J
~,
~~~~~~
{[}\cD_{a},\cD_b{]}
=
-4\,S^2\,\cM_{ab}
~.
\label{3_0-alg-AdS-2}
\eea
\esubeq
As shown in \cite{KLT-M-2011}, in order to define important off-shell supermultiplets and matter 
couplings
in  $\cN=3$ conformal supergravity,  
it is useful to introduce a new basis for the spinor covariant derivatives, $ \cD_\a^I \to \cD_\a^{ij}$,
defined by the rule that 
any SO(3) isovector index is replaced by a  pair of symmetric SU(2) isospinor
indices. Specifically, 
the new covariant derivative $\cD_\a^{ij}$ is defined as\footnote{We refer the reader to section
5 and Appendix A of \cite{KLT-M-2011} for details on our $\cN=3$ isospinor notations including
the properties and explicit definition of the $(\t^I)_{ij}$ matrices.}
\bea
\cD_\a^{ij}:=\cD_\a^I(\t_I)^{ij}=\cD_\a^{ji}~,~~
(\cD_\a^{ij})^*=-\cD_{\a ij}=-\ve_{ik}\ve_{jl}\cD_\a^{kl}
~,~~~~
i=\1,\2
~,
~~
\ve^{\1\2}=\ve_{\2\1}=1
~.
~~~~~~
\eea
In isospinor notations the SO(3) generator $\cN_{KL}$ becomes
\bsubeq
\bea
&\cN_{KL}~\to~\cN_{ijkl}:=\cN_{KL}(\t^K)_{ij}(\t^L)_{kl}
=\hf\ve_{jl}\cJ_{ik}+\hf\ve_{ik}\cJ_{jl}~,
\\
&{\big [} {\cJ}{}^{kl},\cD_{\a}^{ij}{\big]} =\ve^{i(k} \cD_{\a}^{ l)j}+\ve^{j(k} \cD_{\a}^{ l)i}
~,
\eea
\esubeq
where $\cJ^{kl}=\cJ^{lk}$ is the SU(2) generator.
In isospinor notations the (3,0) algebra takes the form
\bsubeq 
\label{3_0-alg-isospinor}
\bea
&\{\cD_\a^{ij},\cD_\b^{kl}\}=
-2\ri\ve^{i(k}\ve^{l)j}\cD_{\a\b}
+2\ri\,S\,\ve_{\a\b}\Big(
\ve^{jl}\cJ^{ik}
+\ve^{ik}\cJ^{jl}
\Big)
+4\ri\,S\,\ve^{i(k}\ve^{l)j}\cM_{\a\b}
~,~~~~~~
\label{3_0-alg-isospinor-1}
\\
%%%%%%
&{[}\cD_{a},\cD_\b^{jk}{]}
=
\,S\,(\g_a)_\b{}^{\g}\cD_{\g}^{jk}
~,~~~~~~
{[}\cD_a,\cD_b{]}=\,-\,4\,S^2\,\cM_{ab}
~.
\label{3_0-alg-isospinor-2}
\eea
\esubeq

\subsubsection{(2,1) AdS superspace}

In the diagonal frame, the (2,1) algebra is
\bsubeq
\bea
&\{\cD_\a^\OI,\cD_\b^\OJ\}=
2\ri\d^{\OI\OJ}\cD_{\a\b}
-4\ri \,S\,\d^{\OI\OJ}\cM_{\a\b}
-4\ri\,S\,\ve_{\a\b}\cN^{\OI\OJ}
~,
\label{2_1-0-1}
\\
%%%%%%%%%%%%
&\{\cD_\a^{\bf 3},\cD_\b^{\bf 3}\}=
2\ri\cD_{\a\b}
+4\ri \,S\, \cM_{\a\b}
~,~~~~~~
\{\cD_\a^\OI,\cD_\b^{\bf 3}\}=0
~,
\label{2_1-0-2}
\\
%%%%%%%%%%%%
&{[}\cD_{a},\cD_\b^\OJ{]}
=
\,S\,(\g_a)_\b{}^\g\cD_{\g}^\OJ
~,
~~~~~~~
{[}\cD_{a},\cD_\b^{\bf 3}{]}
=
-\,S\,(\g_a)_\b{}^\g\cD_{\g}^{\bf 3}
~,
\label{2_1-0-3}
\\
%%%%%%%%%%%%
&
{[}\cD_{a},\cD_b{]}
=
-\,4\,S^2\,\cM_{ab}
~.~~~~~~~~~
\label{2_1-0-4}
\eea
\esubeq
We want to rewrite the previous algebra in isospinor notations. To do that, we first observe that
the algebra is constructed from the AdS algebra in the general frame 
(\ref{alg-AdS-1})--(\ref{alg-AdS-2}) with the choice 
$S^{IJ}=\,S\,\big(\d^{IJ}-(w_3)^I(w_3)^J\big)$ with the vector 
$(w_3)^I=(0,0,\sqrt{2})$ 
in the third direction. It is clear that with an SO(3) rotation 
we can move to a general frame
where  
$S^{IJ}=\,S\,(\d^{IJ}-w^Iw^J)$ and $w^I$ such that $w^Iw_I=2$.
Clearly, the structure group is still SO(2) since, for example, the algebra admits a central extension
with constant central charge field strength given by\footnote{See \cite{KLT-M-2011}
 for the description of 
$\cN$-extended vector multiplets coupled to conformal supergravity. The same analysis holds
for the AdS geometries.}
$w^{IJ}=\ve^{IJK}w_K$, $\cD_A w_K=0$.
The algebra (\ref{2_1-0-1})--(\ref{2_1-0-4}) can be seen to become 
\bsubeq
\bea
\{\cD_\a^I,\cD_\b^J\}&=&
2\ri\d^{IJ}\cD_{\a\b}
-4\ri \,S\,(\d^{IJ}-w^Iw^J)\cM_{\a\b}
-\ri \,S\,\ve_{\a\b}w^{IJ}w^{KL}\cN_{KL}
~,~~~~~~
\\
%%%%%%%%%%%%
{[}\cD_{a},\cD_\b^J{]}
&=&
\,S\,(\d^{J}_{K}-w^Jw_K)(\g_a)_\b{}^\g\cD_{\g}^K
~,
\\
%%%%%%%%%%%%
{[}\cD_{a},\cD_b{]}
&=&
-4\,S^2\,\cM_{ab}
~.
~~~~~~~~~
w^Iw_I=2~,~~
w^{IJ}:=\ve^{IJK}w_K
~.
\eea
\esubeq
From this form it is easy to move to isospinor notations in a general frame.
We obtain
\bsubeq \label{2_1-alg-isospinor}
\bea
\{\cD_\a^{ij},\cD_\b^{kl}\}&=&
-2\ri\ve^{i(k}\ve^{l)j}\cD_{\a\b}
+4\ri \,S\,(\ve^{i(k}\ve^{l)j}+w^{ij}w^{kl})\cM_{\a\b}
\non\\
&&
+\ri \,S\,\ve_{\a\b}\big(\ve^{i(k}w^{l)j}
+\ve^{j(k}w^{l)i}\big)
w^{pq}\cJ_{pq}
~,
\label{2_1-alg-isospinor-1}
\\
%%%%%%%%%%%%
{[}\cD_{a},\cD_\b^{ij}{]}
&=&
\,S\,(\g_a)_\b{}^\g\cD_{\g}^{ij}
-\,S\,w^{ij}w_{kl}(\g_a)_\b{}^\g\cD_{\g}^{kl}
~,
\label{2_1-alg-isospinor-2}
\\
%%%%%%%%%%%%
{[}\cD_{a},\cD_b{]}
&=&
-4\,S^2\,\cM_{ab}
~,
~~~~~~~~~
w^{kl}w_{kl}=2~.
\label{2_1-alg-isospinor-3}
\eea
\esubeq
Note that in the algebra the $R$-symmetry group is generated by the U(1) operator
$w^{pq}\cJ_{pq}$.

%%%%%%%%%%%%%%%%%%%%%%%%%%%%%%%%%%%%%%%%%%%%%%%%

\subsection{$\cN=4$}
\label{AdS-N=4}

In the $\cN=4$ case we have three different AdS geometries: (4,0); (3,1); (2,2).

\subsubsection{(4,0) AdS superspace}
We start with the (4,0) case. This is particularly interesting being the first geometry where 
the covariantly constant $X^{IJKL}$ curvature can be used to deform the AdS geometry.
Since $X^{IJKL}=X\ve^{IJKL}$ for $\cN=4$, 
the (4,0) algebra is
\bsubeq
\bea
&\{\cD_\a^I,\cD_\b^J\}=
2\ri\d^{IJ}\cD_{\a\b}
-4\ri \,S\, \d^{IJ}\cM_{\a\b}
+\ri\ve_{\a\b}\Big(
X\ve^{IJKL}\cN_{KL}
-4\,S\,\cN^{IJ}
\Big)
~,
\label{4_0-alg-AdS-1}
\\
%%%%%%%%%%%%
&{[}\cD_{a},\cD_\b^J{]}
=
\,S\,(\g_a)_\b{}^\g\cD_{\g}^J
~,~~~~~~
{[}\cD_{a},\cD_b{]}
=
-4\,S^2\,\cM_{ab}
~.
\label{4_0-alg-AdS-2}
\eea
\esubeq
Note that the scalar $X$  is a free parameter that does not affect the curvature of the body of AdS.
In particular, we can freely add it also to the $\cN=4$ Minkowski superspace.
Its role is to deform the SO(4) part of the structure group.
To see in details how the $X$ field affects the algebra we change notations for the 
SO(4) isovector indices  and move to pairs of SU(2)$_\rL\times$SU(2)$_\rR$
 isospinor indices
making use of the isomorphism 
 ${\rm SO}(4) \cong  \big( {\rm SU}(2)_{\rL}\times {\rm SU}(2)_{\rR}\big)/{\mathbb Z}_2$.
We define new covariant derivatives $\cD_\a^{i\bai}$ as\footnote{We refer the reader to section
6 and Appendix A of \cite{KLT-M-2011} for details on our $\cN=4$ isospinor notations including
the properties and explicit definition of the $(\t^I)_{i\bai}$ matrices.}
\bea
\cD_\a^{i\bai}:=\cD_\a^I(\t_I)^{i\bai}~,~~~~
(\cD_\a^{i\bai})^*=-\cD_{\a i\bai}=-\ve_{ij}\ve_{\bai\baj}\cD_\a^{j\baj}
~.
\eea
The SO(4) generator $\cN_{KL}$ in isospinor notation takes the form
\bsubeq
\bea
&\cN_{KL}~\to~\cN_{k\bak l\bal}:=\cN_{KL}(\t^K)_{k\bak}(\t^L)_{l\bal}
=\ve_{\bak\bal}\bL_{kl}+\ve_{kl}\bR_{\bak\bal}~,
\\
&
{\big [} {\bL}{}^{kl},\cD_{\a}^{i\bai}{\big]} =\ve^{i(k} \cD_{\a}^{ l)\bai}
~,~~~
{\big [} {\bR}{}^{\bak\bal},\cD_{\a}^{i\bai}{\big]} =\ve^{\bai(\bak} \cD_{\a}^{ i\bal)}~,
\eea
\esubeq
where $\bL_{kl}$ and $\bR_{\bak\bal}$ are respectively the left and right SU(2) generators.
Finally, the (4,0) algebra becomes
\bsubeq \label{4200}
\bea
\{\cD_\a^{i\bai},\cD_\b^{j\baj }\}&=&
2\ri\ve^{ij}\ve^{\bai \baj }\cD_{\a\b}
+{2\ri}\ve_{\a\b}\ve^{\bai \baj }(2S+X)\bL^{ij}
+2\ri\ve_{\a\b}\ve^{ij}(2S-X)\bR^{\bai \baj }
\non\\
&&
-4\ri \,S\,\ve^{ij}\ve^{\bai \baj }\cM_{\a\b}
~,
\\
%%%%%%%%%%%%
{[}\cD_a,\cD_\b^{j\baj}{]}&=&
\,S\,(\g_a)_\b{}^\g\cD_\g^{j\baj}
~,~~~~~~
{[}\cD_a,\cD_b{]}=\,-\,4\,S^2\,\cM_{ab}
~.
\eea
\esubeq
It is interesting to note that for generic value of $X$ the 
entire SO(4) group has non-trivial curvature in the algebra.
But there are two points in which either the SU(2)$_\rR$ or the SU(2)$_\rL$ curvatures are zero
and the structure group is reduced.
These are given by
\bea
X=\pm \,2\,S\,
~.
\label{4.21}
\eea

\subsubsection{(2,2) AdS superspace}

The next case we consider is the  (2,2) geometry.
In the diagonal frame this takes exactly the form (\ref{alg-AdS-2-1})--(\ref{alg-AdS-2-2}) 
where the $\cN^{\OI\OJ}$ rotates the directions $I={\bf 1},{\bf 2}$ and the 
$\cN^{\UI\UJ}$ rotates the directions $I={\bf 3},{\bf 4}$ in the isovector space.
The torsion $S^{IJ}=\,S\,{\rm diag}({1,1,-1,-1})$ is traceless $\d_{IJ}S^{IJ}=0$.
We can use this information to 	derive the (2,2) geometry in isospinor notations. 
The traceless condition tells us that 
\bea
S^{IJ}~\to~ (\t_I)^{i\bai}(\t_J)^{j\baj}S^{IJ}=\cS^{ij\bai\baj}=\cS^{ji\bai\baj}=\cS^{ij\baj\bai}
~,
\eea
which can be easily seen by remembering that \cite{KLT-M-2011}
\bea
\d^{IJ}~\to~ (\t_I)^{i\bai}(\t_J)^{j\baj}\d^{IJ}=\ve^{ij}\ve^{\bai\baj}
~.
\eea
The constraint (\ref{alg-constr-SS}) in isospinor notation gives the condition
\bea
\cS^{ij\bai\baj}=   S\,l^{ij}r^{\bai\baj}
~,~~~~~~
l^{kl}l_{kl}=r^{\bak\bal}r_{\bak\bal}=2
~.
\eea
In a general frame, in isospinor notations, the (2,2) algebra then takes the 
following form
\bsubeq
\bea
&\{\cD_\a^{i\bai},\cD_\b^{j\baj }\}=
2\ri\ve^{ij}\ve^{\bai \baj }\cD_{\a\b}
-2\ri\,S\,\ve_{\a\b}\ve^{ij}\,r^{\bai\baj}\bL
-2\ri\,S\,\ve_{\a\b}\ve^{\bai \baj }\,l^{ij}\bR
-4\ri\,S\, l^{ij}\,r^{\bai \baj }\cM_{\a\b}
~,~~~~~~~~~
\\
%%%%%%%%%%%%
&{[}\cD_a,\cD_\b^{j\baj}{]}
=\,S\,l^j{}_k \,r^\baj{}_\bak (\g_a)_\b{}^\g\cD_\g^{k\bak}
~,~~~~~~
{[}\cD_a,\cD_b{]}=\,-\,4\,S^2\,\cM_{ab}~,
\eea
\esubeq
where we have defined the U(1)$_\rL$ and U(1)$_{\rR}$ generators
\bea
\bL:=l^{kl}\bL_{kl}~,~~~~~~
\bR:=r^{\bak\bal}\bR_{\bak\bal}
~.
\eea

\subsubsection{(3,1) AdS superspace}

We are left with the (3,1) case. In the diagonal frame the geometry is
\bsubeq
\bea
&\{\cD_\a^\OI,\cD_\b^\OJ\}=
2\ri\d^{\OI\OJ}\cD_{\a\b}
-4\ri\,S\,\d^{\OI\OJ}\cM_{\a\b}
-4\ri\,S\,\ve_{\a\b}\cN^{\OI\OJ}
~,
\label{3_1-alg-AdS-0-1a}
\\
%%%%%%%%%%%%
&\{\cD_\a^{\bf 4},\cD_\b^{\bf 4}\}=
2\ri\cD_{\a\b}
+4\ri \,S\, \cM_{\a\b}
~,~~~~~~
\{\cD_\a^\OI,\cD_\b^{\bf 4}\}=
\{\cD_\a^{\bf 4},\cD_\b^\OJ\}=
0
~,
\label{3_1-alg-AdS-0-1c}
\\
%%%%%%%%%%%%
&{[}\cD_{a},\cD_\b^\OJ{]}
=
\,S\,(\g_a)_\b{}^\g\cD_{\g}^\OJ
~,
~~~~~~
{[}\cD_{a},\cD_\b^{\bf 4}{]}
=
-\,S\,(\g_a)_\b{}^\g\cD_{\g}^{\bf 4}
~,
\label{3_1-alg-AdS-0-3/2}
\\
%%%%%%%%%%%%
&{[}\cD_{a},\cD_b{]}=
-\,4\,S^2\,\cM_{ab}
~,
\label{3_1-alg-AdS-0-2}
\eea
\esubeq
where here $\cN^{\OI\OJ}$ generate SO(3) rotations of the $I={\bf 1},{\bf 2},{\bf 3}$ isovector 
indices leaving invariant the $(w_4)^I=(0,0,0,\sqrt{2})$ vector.
Similarly to the (2,1) case, with a SO(4) rotation we can rewrite the (3,1) geometry in a general 
frame as
\bsubeq
\bea
&\{\cD_\a^I,\cD_\b^J\}=
2\ri\d^{IJ}\cD_{\a\b}
-4\ri \,S\,\big(\d^{IJ}-w^Iw^J\big)\cM_{\a\b}
-4\ri\ve_{\a\b}
\,S\,
\hat{\cN}^{IJ}
~,
\label{3_1-alg-AdS-1-1}
\\
%%%%%%%%%%%%
&{[}\cD_{a},\cD_\b^J{]}
=
\,S\,\big(\d^{J}_{K}-w^Jw_K\big)(\g_a)_\b{}^\g\cD_{\g}^K
~,
\label{3_1-alg-AdS-1-3/2}
\\
%%%%%%%%%%%%
&{[}\cD_{a},\cD_b{]}
=
-\,4\,S^2\,\cM_{ab}
~,
~~~~~~
\hat{\cN}^{IJ}
:=\big(\d^{K[I}-w^Kw^{[I}\big)\cN_K{}^{J]}
~,
\label{3_1-alg-AdS-1-2}
\eea
\esubeq
with $w^{I}$ satisfying  $w^{I}w_I=2$ but otherwise an arbitrary isovector.
The operator $\hat{\cN}^{IJ}$ 
generates an SO(3) algebra inside SO(4). This can be easily seen by 
observing that $w^I$ is left invariant, $\hat{\cN}^{KL}w^I=0$ and then $\hat{\cN}^{IJ}$
generates rotations orthogonal to $w^I$.
Note that the previous representation of the (3,1) algebra is the same in describing the general
($\cN-1$,1) cases.
By using (\ref{3_1-alg-AdS-1-1})--(\ref{3_1-alg-AdS-1-2})
we can derive a representation of the (3,1) algebra in isospinor notations.
This takes the form
\bsubeq
\bea
\{\cD_\a^{i\bai},\cD_\b^{j\baj}\}&=&
2\ri\ve^{ij}\ve^{\bai\baj}\cD_{\a\b}
+2\ri\ve_{\a\b}\,S\,\Big(
\ve^{\bai\baj}\big(\bL^{ij}
+w^{i}{}_{\bak}w^{j}{}_{\bal}\bR^{\bak\bal}\big)
+\ve^{ij}\big(
\bR^{\bai\baj}
+w_k{}^{\bai}w_{l}{}^{\baj}\bL^{kl}\big)
\Big)
\non\\
&&
-4\ri \,S\,\big(\ve^{ij}\ve^{\bai\baj}-w^{i\bai}w^{j\baj}\big)\cM_{\a\b}
~,
\label{3_1-alg-AdS-s-1}
\\
%%%%%%%%%%%%
{[}\cD_{a},\cD_\b^{j\baj}{]}
&=&
\,S\,\big(\d^j_k\d^\baj_\bak-w^{j\baj}w_{k\bak}\big)(\g_a)_\b{}^\g\cD_{\g}^{k\bak}
~,~~~~~~
\\
%%%%%%%%%%%%
{[}\cD_{a},\cD_b{]}
&=&
-\,4\,S^2\,\cM_{ab}
~,~~~~
w^{k\bak}w_{k\bak}=2~,~
w^{k\bak}w_{j\bak}=\d^k_j~,~
w^{k\bak}w_{k\baj}=\d^\bak_\baj
~.~~~~~~
\label{3_1-alg-AdS-s-2}
\eea
\esubeq
Note that the spinor covariant derivatives algebra can be rewritten as
\bea
\{\cD_\a^{i\bai},\cD_\b^{j\baj}\}&=&
2\ri\ve^{ij}\ve^{\bai\baj}\cD_{\a\b}
+2\ri\ve_{\a\b}\,S\,\Big(
\ve^{\bai\baj}\d^i_k\d^j_l
+\ve^{ij}w_k{}^{\bai}w_{l}{}^{\baj}
\Big)\cJ^{kl}
\non\\
&&
-4\ri \,S\,\big(\ve^{ij}\ve^{\bai\baj}-w^{i\bai}w^{j\baj}\big)\cM_{\a\b}
~,
\label{3_1-alg-AdS-s2-1}
\eea
or equivalently as
\bea
\{\cD_\a^{i\bai},\cD_\b^{j\baj}\}&=&
2\ri\ve^{ij}\ve^{\bai\baj}\cD_{\a\b}
+2\ri\ve_{\a\b}\,S\,\Big(
\ve^{ij}\d^\bai_\bak\d^\baj_\bal
+\ve^{\bai\baj}w^i{}_{\bak}w^{j}{}_{\bal}
\Big)
\cJ^{\bak\bal}
\non\\
&&
-4\ri \,S\,\big(\ve^{ij}\ve^{\bai\baj}-w^{i\bai}w^{j\baj}\big)\cM_{\a\b}
~,
\label{3_1-alg-AdS-s3-1}
\eea
where we have defined
\bsubeq
\bea
\cJ^{kl}&:=&\big(\bL^{kl}+w^{k}{}_{\bak}w^{l}{}_{\bal}\bR^{\bak\bal}\big)
~,
\\
\cJ^{\bak\bal}&:=&
\big(\bR^{\bak\bal}+w_{k}{}^{\bak}w_{l}{}^{\bal}\bL^{kl}\big)
~.
\eea
\esubeq
The operator $\cJ^{kl}=w^k{}_\bak w^l{}_\bal\cJ^{\bak\bal}$,
or equivalently $\cJ^{\bak\bal}=w_k{}^\bak w_l{}^\bal\cJ^{kl}$,
generates the residual SU(2) algebra of the (3,1) AdS geometry
and leaves $w^{i\bai}$ invariant.

%%%%%%%%%%%%%%%%%%%%%%%%%%%%%%%%%%%%%%%%%%%%%%%
%%%%%%%%%%%%%%%%%%%%%%%%%%%%%%%%%%%%%%%%%%%%%%%
%%%%%%%%%%%%%%%%%%%%%%%%%%%%%%%%%%%%%%%%%%%%%%%

\section{Rigid $\cN=3$  supersymmetric field theories in AdS: Off-shell multiplets and invariant 
actions}
\setcounter{equation}{0}

In this and the next sections, our goal is to apply the supergravity techniques of  \cite{KLT-M-2011} 
to describe general 
nonlinear $\s$-models in $\rm AdS_3$ possessing $\cN=3$ supersymmetry. We recall that the 
case of $\cN=2$ 
AdS supersymmetry has already been studied in \cite{KT-M-2011}.  Similar in some aspects 
to $\cN=3$, 
the case of $\cN=4$ AdS supersymmetry nevertheless requires a separate analysis that will be 
given elsewhere. 

In discussing off-shell supermultiplets and supersymmetric actions, we first  give a unified 
presentation 
that applies equally well to the (3,0) and (2,1) AdS supersymmetry types. After that, 
we spell out those technical aspects of $\cN=3$ supersymmetric theories in $\rm AdS_3$
which look essentially different for the cases (3,0) and (2,1). 

For our subsequent consideration, it is useful to rewrite the (anti) commutation relations 
for the (3,0) and (2,1) covariant derivatives 
in a unified form (which is inspired by the algebra of covariant derivatives in 
$\cN=3$ conformal supergravity \cite{KT-M-2011}):
\begin{subequations}
\bea
\{\cD_\a^{ij},\cD_\b^{kl}\}&=&
-2\ri\ve^{i(k}\ve^{l)j}\cD_{\a\b}
-4\ri(\cS^{ijkl}-\ve^{i(k}\ve^{l)j}\cS)\cM_{\a\b} \non \\
&&
-\ri\ve_{\a\b}(\ve^{jl}\cS^{ikpq}
+\ve^{ik}\cS^{jlpq})\cJ_{pq}
+2\ri\ve_{\a\b}\cS\Big(
\ve^{jl}\cJ^{ik}
+\ve^{ik}\cJ^{jl}
\Big) ~,
\label{alg-3331}\\
{[}\cD_{\a\b},\cD_\g^{ij}{]}
&=&
-2\cS^{ijkl}\ve_{\g(\a}\cD_{\b)kl}
-2\cS\ve_{\g(\a}\cD_{\b)}^{ij}~.
\label{alg-3332}
\eea
\end{subequations}
In (\ref{alg-3331})--(\ref{2_1-Killing_iso_3-0}), 
the covariantly constant tensors $\cS^{ijkl} = \cS^{(ijkl)}$ and $\cS$ have the following 
explicit expressions for the (3,0) and (2,1) AdS superspaces
\bea
\mbox{(3,0) AdS}:&&~~~~~~
\cS=S~,\quad 
\,~\cS^{ijkl}=0~;
\label{S-30}
\\
\mbox{(2,1) AdS}:&&~~~~~~
\cS=\frac{1}{3}S~,\quad
\cS^{ijkl}=-Sw^{(ij}w^{kl)}
~,
\label{S-21}
\eea
where the covariantly constant tensor $w^{ij}= w^{(ij)}$ is normalised by  $w^{ij}w_{ij}=2$.

The $\cN=3$ Killing equations 
\bea
\big[ \x +\hf \L^{ab}\cM_{ab} +\hf \L^{ij} \cJ_{ij}, \cD_C \big] =0 ~, \qquad
\x=\x^a\cD_a+\x^\a_{ij}\cD_\a^{ij}
\label{4.4}
\eea
are equivalent to
\bsubeq
\bea
\cD_\a^{ij}\x_\b^{kl}
&=&
\hf\ve_{\a\b}\big(\ve^{ik}\L^{jl}
+\ve^{jl}\L^{ik}\big)
+(\cS^{ijkl}-\ve^{i(k}\ve^{l)j}\cS)\x_{\a\b}
-\hf\ve^{i(k}\ve^{l)j}\L_{\a\b}
~,
\label{2_1-Killing_iso_1-0}
\\
%%%%%%
0&=&
\cD_{ \g}^{ij}\x^{\a\g}
+6\ri\x^{\a ij}
~,~~~
0=
\cD_\g^{ij}\L^{\a\g}
+12\ri \x^{\a}_{kl}(\cS^{ijkl}-\ve^{i(k}\ve^{l)j}\cS)
\label{2_1-Killing_iso_2-0}
\\
%%%%%%
0&=&
\cD_{(\a}^{ ij}\x_{\b\g)}
=\cD_{(\a}^{ ij}\L_{\b\g)}
~.
\label{2_1-Killing_iso_3-0}
\eea
\esubeq
These relations imply, in particular, 
the following equations
\bea
\cD_{\a kl}\L^{kl}=0
~,~~~
\cD_\a{}^{(i}{}_k\L^{j)k}
=
-2\ri\Big(
4\cS\x_{\a}^{ij}
+\cS^{ijkl}\x_{\a kl}
\Big)
~,~~~
\cD_\a^{(ij}\L^{kl)}
=
-4\ri \x_{\a}{}^{(i}{}_p\cS^{jkl)p}
\label{4.6}
\eea
which will be important for our subsequent consideration. 
We recall that the parameter  $\L^{ij}$ is real,   $\overline{\L^{ij}}=\L_{ij}$. 

In  the (3,0) and (2,1) cases, 
the $R$-symmetry groups are SU(2) and U(1) respectively. 
In the case of (2,1) AdS supersymmetry,    
the parameter  $\L^{ij}$ has the form
\bea
\mbox{(2,1) AdS}:&&~~~~~~\L^{ij}=w^{ij}\L~,~~~~~~
\L=\overline{\L}
~.
\eea

\subsection{Covariant projective supermultiplets}
In complete analogy with matter couplings in $\cN=3$ supergravity \cite{KLT-M-2011}, 
a large class of rigid supersymmetric theories in (3,0) and (2,1) AdS superspaces can be 
formulated in terms of covariant projective supermultiplets.
Before introducing these supermultiplet, a few words are in order regarding the so-called 
projective superspace approach.

The projective superspace approach \cite{KLR,LR1,LR2} is a method to construct 
off-shell 4D $\cN=2$ super-Poincar\'e invariant theories in the superspace 
${\mathbb R}^{4|8} \times {\mathbb C}P^1$ introduced 
for the first time by Rosly \cite{Rosly}.\footnote{The same superspace is used 
within the harmonic superspace approach \cite{GIKOS,GIOS} which is more general than 
the projective one but less useful for various $\s$-model applications. 
The precise relationship between the harmonic and projective superspace
formulations is spelled out in  \cite{K98}.}
The most important projective supermultiplets are: the $\cO(1)$ multiplet \cite{Rosly}
(equivalent to the on-shell hypermultiplet \cite{Sohnius}); 
the real $\cO(2) $ multiplet \cite{KLR} (equivalent to the $\cN=2$ tensor multiplet 
\cite{SSW}); 
the $\cO(n)$ multiplets \cite{KLT,LR1}, where $n=3,4,\dots $; 
the polar (arctic + antarctic) multiplet \cite{LR1}; the tropical multiplet \cite{LR2}.
These multiplets  are off-shell except the $\cO(1) $ multiplet. 
The projective superspace approach  was extended to conformal supersymmetry
\cite{K-comp,K-conf} and supergravity \cite{KT-M_5D,KT-M_5Dconf},  more than  
twenty years after 
the original publication on self-interacting $\cN=2$ tensor multiplets \cite{KLR}. 
The original 5D $\cN=1$ supergravity construction of  \cite{KT-M_5D,KT-M_5Dconf} 
has successfully been extended to 4D $\cN=2$ supergravity 
\cite{KLRT-M_4D-1,KLRT-M_4D-2},  3D $\cN=3$ and $\cN=4$ supergravity theories 
\cite{KT-M-2011}, 2D $\cN= (4,4) $ supergravity \cite {GTM_2D_SUGRA},  and most recently 
6D $\cN=(1,0)$ supergravity \cite{LT-M-2012}. 

 A {\em covariant projective supermultiplet} of weight $n$,
$Q^{(n)}(z^M,v^i)$, is defined 
to be a Lorentz-scalar superfield that 
lives on the appropriate $\cN=3$ AdS superspace $\cM^{3|6}$
 (which is  $\rm AdS_{(3|3,0)}$ or  $\rm AdS_{(3|2,1)}$),
is holomorphic with respect to isospinor variables $v^i $ on an open domain of 
${\mathbb C}^2 \setminus  \{0\}$, 
and is characterised by the following conditions:

(i) $Q^{(n)}$  is  a homogeneous function of $v$ 
of degree $n$, that is,  
\be
Q^{(n)}(z,c\,v)\,=\,c^n\,Q^{(n)}(z,v)~, \qquad c\in \mathbb{C}^* \equiv {\mathbb C} \setminus  \{0\}~;
\label{weight}
\ee

(ii)  Under the appropriate AdS isometry supergroup, 
which  is  ${\rm OSp(}3|2;{\mathbb R}) \times {\rm Sp}(2,{\mathbb R})$
or ${\rm OSp(}2|2;{\mathbb R}) \times {\rm OSp}(1|2;{\mathbb R})$, $Q^{(n)}$ 
transforms as follows:
\bea
\d_\x Q^{(n)} 
&=& \Big( \x + \hf \L^{ij} \cJ_{ij} \Big) Q^{(n)} ~,  
\non \\ 
\L^{ij} \cJ_{ij}  Q^{(n)}&=& -\Big(\L^{(2)} {\bm \pa}^{(-2)} 
-n \, \L^{(0)}\Big) Q^{(n)} ~, \qquad {\bm \pa}^{(-2)} :=\frac{1}{(v,u)}u^{i}\frac{\pa}{\pa v^{i}}~.
\label{harmult1}   
\eea 
where $\x$ denotes an arbitrary  AdS Killing vector field, eq.  \eqref{2.28},
 and $\L^{ij}$ the associated SU(2) parameter defined by \eqref{2.29}.\footnote{In the case of (2,1) 
 AdS supersymmetry, 
the  parameter  $\L^{ij}$ is constrained to be  $\L^{ij}= \L w^{ij}$, which corresponds to an SO(2) 
subgroup of SU(2).} 
In eq. \eqref{harmult1},   
we have introduced 
\bea
\L^{(2)} :=\L^{ij}\, v_i v_j 
~,\qquad
\L^{(0)} :=\frac{v_i u_j }{(v,u)}\L^{ij}~,
\qquad (v,u):=v^iu_i~.
\label{W2t3}
\eea
The transformation law \eqref{harmult1} involves 
an additional  isotwistor  $u_{i}$, which is 
only subject to the condition $(v,u)\ne0$, and  otherwise is completely arbitrary.
Both $Q^{(n)}$ and $\d_\x Q^{(n)}$ are independent of $u_i$. 

(iii) $Q^{(n)}$  obeys the analyticity constraint
\be
\cD^{(2)}_{\a} Q^{(n)}  =0~, \qquad \cD_\a^{(2)}:=v_iv_j\cD_\a^{ij}
~.
\label{ana}
\ee  

The analyticity constraint (\ref{ana}) and the homogeneity condition (\ref{weight}) 
are consistent with the interpretation that 
the isospinor
$ v^{i} \in {\mathbb C}^2 \setminus\{0\}$ is   defined modulo the equivalence relation
$ v^{i} \sim c\,v^{i}$,  with $c\in {\mathbb C}^*$, {hence it parametrizes ${\mathbb C}P^1$}.
Therefore, the projective multiplets live in ${\cM}^{3|6} \times {\mathbb C}P^1$.

Two comments are in order. Firstly, 
 it  follows from  eq. (\ref{harmult1}) that
\bea
\cJ^{(2)} Q^{(n)}=0~,\qquad
\cJ^{(2)}:=v_iv_j\cJ^{ij}~.
\label{J++}
\eea
Secondly, the constraints (\ref{ana}) are fully consistent due to 
the facts that $Q^{(n)}$ is a Lorentz scalar, and
the operators $\cD^{(2)}_{\a}$ obey the anti-commutation relations 
\bea
&\{\cD_\a^{(2)},\cD_\b^{(2)}\}=
-4\ri \cS^{(4)}\cM_{\a\b}
~,
\label{N=3D2D2}
\eea
with
\bea
&
\cS^{(4)}:=v_iv_jv_kv_l\cS^{ijkl}
~.
\label{N=3Ta-1}
\eea

A more general family of off-shell supermultiplets is obtained by removing the condition (iii) 
in the above definition, while keeping intact the conditions (i) and (ii). Such supermultiplets are 
called {\it isotwistor}.  These superfields can be used to construct projective ones with the aid 
of the so-called analytic projection operator
\bea
\D^{(4)}:=\frac{\ri}{4}\Big(\cD^{(4)}-4\ri\cS^{(4)}\Big)~, \qquad 
\cD^{(4)}:=\cD^{(2)\a}\cD^{(2)}_\a~.
\eea
If  $U^{(n-4)}(z,v)$ is an isotwistor superfield, then $Q^{(n)} (z,v):= \D^{(4)} U^{(n-4)} (z,v)$ is 
a covariant projective superfield, 
\bea 
\cD^{(2)}_\a \D^{(4)} U^{(n-4)} =0~.
\eea

There exists 
a real structure on the space of projective multiplets 
known as the smile conjugation.\footnote{The   smile conjugation was
pioneered by Rosly \cite{Rosly} and re-discovered in \cite{GIKOS,KLR}.}
Given a  weight-$n$ projective multiplet $ Q^{(n)} (v^{i})$, 
its {\it smile conjugate},
$ \breve{Q}^{(n)} (v^{i})$, is defined by 
\bea
 Q^{(n)}(v^{i}) \longrightarrow  {\bar Q}^{(n)} ({\bar v}_i) 
  \longrightarrow  {\bar Q}^{(n)} \big({\bar v}_i \to -v_i  \big) =:\breve{Q}^{(n)}(v^{i})~,
\label{smile-iso}
\eea
with ${\bar Q}^{(n)} ({\bar v}_i)  :=\overline{ Q^{(n)}(v^{i} )}$
the complex conjugate of  $ Q^{(n)} (v^{i})$, and ${\bar v}_i$ the complex conjugate of 
$v^{i}$. One can show that $ \breve{Q}^{(n)} (v)$ is a weight-$n$ projective multiplet.
In particular,   $ \breve{Q}^{(n)} (v)$
obeys the analyticity constraint $\cD_\a^{(2)}\breve{Q}^{(n)} =0$,
unlike the complex conjugate of $Q^{(n)}(v) $.
One can also check that 
\bea
\breve{ \breve{Q}}^{(n)}(v) =(-1)^n {Q}^{(n)}(v)~.
\label{smile-iso2}
\eea
Therefore, if  $n$ is even, one can define real projective multiplets, 
 $\breve{Q}^{(2n)} = {Q}^{(2n)}$.
Note that geometrically, the smile-conjugation is complex conjugation composed
with the antipodal map on the projective space ${\mathbb C}P^1$.

We now list several projective multiplets that can be  used to describe superfield 
dynamical variables.
A complex $\cO(m) $ multiplet, 
with $m=1,2,\dots$,   is described by a  weight-$m$ projective superfield $H^{(m)} (v)$ 
of the form:
\bea
H^{(m)} (v) &=& H^{i_1 \dots i_{m}} v_{i_1} \dots v_{i_{m}} 
~.
\eea
The analyticity constraint (\ref{ana}) is equivalent to 
\bea
\cD_\a^{(ij} H^{k_1 \dots k_{m} )} =0~.
\eea
If $m$ is even, $m=2n$, we can define a real $\cO(2n) $ multiplet\footnote{In 4D $\cN=2$ 
Poincar\'e supersymmetry, the real $\cO(2n)$ multiplets, with $n>1$, 
and their  self-interactions were introduced for the first time by Ketov and Tyutin \cite{KLT} 
and re-discovered in \cite{LR2}.}
obeying 
the reality condition $\breve{H}^{(2n)}  = {H}^{(2n)} $, or equivalently
\bea
\overline{ H^{i_1 \dots i_{2n}} } &=& H_{i_1 \dots i_{2n}}
=\ve_{i_1 j_1} \cdots \ve_{i_{2n} j_{2n} } H^{j_1 \dots j_{2n}} ~.
\eea
The field strength of an Abelian  vector multiplet is a real $\cO(2) $ multiplet  \cite{KLT-M-2011}. 
For $n>1$, 
the real $\cO(2n) $ multiplet can be used to describe an off-shell (neutral) hypermultiplet. 

An off-shell (charged) hypermultiplet can be described in term of the so-called {\it arctic} 
weight-$n$ multiplet $\U^{(n)} (v)$ which is defined to be 
holomorphic  in the north chart  $\mathbb C$, 
of the projective space ${\mathbb C}P^1 ={\mathbb C} \cup \{\infty \}$: 
\bea
\U^{(n)} ( v) &=&  (v^{\1})^n\, \U^{[n]} ( \z) ~, \qquad 
\U^{ [n] } ( \z) = \sum_{k=0}^{\infty} \U_k  \z^k 
~, 
\label{arctic1}
\eea
and  its smile-conjugate {\it antarctic} multiplet $\breve{\U}^{(n)} (v) $,
 \bea
\breve{\U}^{(n)} (v) &=& 
(v^{\2}  \big)^{n}\, \breve{\U}^{[n]}(\z) =
(v^{\1} \,\z \big)^{n}\, \breve{\U}^{[n]}(\z) ~, \qquad
\breve{\U}^{[n]}( \z) = \sum_{k=0}^{\infty}  {\bar \U}_k \,
\frac{(-1)^k}{\z^k}~.~~~
\label{antarctic1}
\eea
Here we have introduced the inhomogeneous complex coordinate 
$\z= v^{\2}/v^{\1}$ on the north chart of  ${\mathbb C}P^1$.
The pair consisting of $\U^{[n]} ( \z)$ and $\breve{\U}^{[n]}(\z) $ 
constitutes the so-called polar weight-$n$ multiplet.

\subsection{Supersymmetric action}

In order to formulate the dynamics of rigid $\cN=3$ supersymmetric field theories in $\rm AdS_3$, 
a manifestly supersymmetric action principle is required. It can be readily constructed by 
restricting the locally supersymmetric action introduced in \cite{KLT-M-2011} to the appropriate 
AdS superspace.  The action is generated by a Lagrangian  $\cL^{(2)} (z,v)$, 
which is a covariant weight-2 real projective multiplet, and has the form: 
\bea
S[\cL^{(2)}]&=&
\frac{1}{2\pi\ri} \oint_\g  (v, \rd v)
\int \rd^3 x \,{\rm d}^6\q\,E\, \cC^{(-4)}\cL^{(2)}~, 
\qquad E^{-1}= {\rm Ber}(E_A{}^M)~.
\label{InvarAc}
\eea
Here the line integral  is carried out over a closed contour
$\g =\{v^i(t)\}$ in ${\mathbb C}P^1$.
The action involves an isotwistor superfield $\cC^{(-4)} (z,v)$ defined by
\bea
\cC^{(-4)}:=\frac{\cU^{(n)}}{\D^{(4)}\cU^{(n)}}
~,
\eea
for some isotwistor multiplet $\cU^{(n)}$ such that 
$1/ \D^{(4) } \cU^{(n)} $ is well defined.  
The superfield $\cC^{(-4)}$ is 
required to write the action as an integral over the full AdS  superspace.
It is actually a purely gauge degree of freedom 
in the sense that  \eqref{InvarAc} is independent of the explicit choice of $\cU^{(n)}  $.
Indeed, varying $\cU^{(n)}$ gives
$$
\d \cC^{(-4)}=\frac{\d \cU^{(n)}}{\D^{(4)}\cU^{(n)}} 
- \frac{\cU^{(n)} \D^{(4)} \d \cU^{(n)}}{(\D^{(4)}\cU^{(n)})^2}~.
$$
In the contribution to $\d S[\cL^{(2)}] $ which comes from the second term, 
we can integrate by parts, to strip $\d \cU^{(n)}$ of $\D^{(4)}$, and make use of the fact that  
$\cL^{(2)}$ and $\D^{(4)} \cU^{(n)} $ are covariant projective multiplets.
As a result, we obtain $\d S[\cL^{(2)}] =0$.

In the case of (2,1) AdS supersymmetry,  there is a simple choice for $\cC^{(-4)}$:
\bea
\cC^{(-4)}=\frac{ 1}{\D^{(4)} 
1} = - \frac{1}{S (w^{(2)})^2 } ~, \qquad 
w^{(2)}:= v_i v_j w^{ij}~.
\eea

%%%%%%%%%%%%%%%%%%%%%%%%%%%%%%%%

\subsection{Supersymmetric action: Integrating out all the fermionic directions}
\label{N3aComp}

The action  \eqref{InvarAc}  is manifestly invariant 
under arbitrary isometry transformations of the appropriate AdS superspace, 
 $\rm AdS_{(3|3,0)}$ or  $\rm AdS_{(3|2,1)}$. The price to pay for this is two-fold:
(i) the action involves  the superfield $\cC^{(-4)}$ which is a purely gauge degree of freedom; 
(ii) the action is given by  an integral over six Grassmann variables while the Lagrangian 
$\cL^{(2)}$ depends
only on four of these coordinates. Both drawbacks can be eliminated, at the cost of losing 
the manifest invariance under the AdS isometry supergroup, if one integrates out {\it two} or 
{\it all} of the six fermionic directions. To achieve this, one could use the powerful method
of normal coordinates around a submanifold of curved superspace \cite{KT-M-normal}.
Here we are going to use an alternative technique which was first developed to derive the 
$\cN=1$ supersymmetric action in  $\rm AdS_5$ \cite{KT-M}.

Our point of departure is  the  $\cN=3$ projective superspace action 
in three-dimensional Minkowski space 
which was introduced in \cite{KPT-MvU-2011}.
It has the form
\bea
S[L^{(2)}]=\frac{1}{8\p} \oint_{\g}  { v_i {\rm d} v^i }
\int {\rm d}^3x \, \big(D^{(-2)}\big)^2 \big(D^{(0)}\big)^2 L^{(2)}  \Big|_{\q=0}
~,
\label{flat-Ac}
\eea
where the Lagrangian $L^{(2)}(z,v)$ is a real weight-two projective multiplet, 
and  the operators $D^{(-2)}_\a$ and $D^{(0)}_\a$ are defined in terms of the {\it flat} spinor 
covariant derivatives $D_\a^{ij} $ as follows 
\bea
D^{(-2)}_\a:=\frac{u_iu_j}{(v,u)^2}D_\a^{ij}~,~~~~~~
D^{(0)}_\a:=\frac{v_iu_j}{(v,u)}D_\a^{ij}
~.
\eea
These operators depend not only on the isotwistor
$v^i (t) $, which varies along the integration contour,  
but  also on a constant ($t$-independent) isotwistor $u_i$ chosen in such a way that 
$v_i(t)$ and $u_i$  
are linearly independent at each point of the contour $\g$,
that is $\big(v(t),u\big) \neq 0$. The action \eqref{flat-Ac} is actually independent of $u_i$, since it 
proves to be invariant under arbitrary
projective transformations of the form
\be
\big(u_i{}\,,\,v_i (t){}\big)~\to~\big(u_i{}\,,\, v_i (t){} \big)\,R(t)~,~~~~~~R(t)\,=\,
\left(\begin{array}{cc}a(t)~&0\\ b(t)~&c(t)\end{array}\right)\,\in\, 
{\rm GL}(2,\mathbb{C})~,
\label{proj-inv}
\ee
where the matrix elements $a(t)$ and $b(t)$ obey the first-order differential equations
\bea
\dt{a}=b\,{(\dt{v}, v)\over (v,u)}~, \qquad \dt{b}=-b\,{(\dt{v}, u)\over (v,u)}~,
\label{ode}
\eea
with $\dt{f}:= \rd f(t)/ \rd t$ for any function $f(t)$. This invariance follows from the following 
properties of the Lagrangian:
(i) $L^{(2)} (v)$ is a homogeneous function of $v^i$ of degree two; and (ii) $L^{(2)} (v)$
obeys the analyticity condition 
\bea
D^{(2)}_\a L^{(2)} (v)=0~, \qquad D^{(2)}_\a:=v_i v_j D_\a^{ij}~.
\eea
It turns out that the property (ii) suffices to prove that  the action  \eqref{flat-Ac} is 
invariant under the standard $\cN=3$ super-Poincar\'e transformations in three dimensions 
\cite{KPT-MvU-2011}.

We now try to generalise the above construction to the AdS case. 
Let $z^M = (x^m, \q^\m_{\imath \jmath})$
be local coordinates of the AdS superspace.
Given a  tensor superfield $U(x,\q)$,  
we define its restriction to the body of the superspace, $ \q^\m_{\imath \jmath}=0$, 
specifically\footnote{In what follows, we will also introduce a single bar-projection, $U|$, 
to be the restriction of $U$ to a certain $\cN=2$ subspace of the  $\cN=3$ AdS superspace under 
consideration.} 
\be
U||:= U(x,\q )|_{\q_{\imath \jmath}=0}~.
\label{projection-0}
\ee
We also define the double-bar projection of the covariant derivatives 
\be
\cD_{{A}} ||:= 
E_{A}{}^M||\pa_M
+\hf\O_A{}^{bc}||\cM_{bc}
+\hf\F_A{}^{kl}||\cJ_{kl}
~.
\ee
Since for both the (3,0) and (2,1) AdS geometries 
it holds that ${[}\cD_{a},\cD_b{]} =-4\,S^2\,\cM_{ab}$, 
we can use the freedom to perform general coordinate and local structure group transformations 
to choose a (Wess-Zumino) gauge in which
\be
\cD_{{a}} || = \nabla_{ a} = e_{a}{}^{m} (x) \, \pa_{ m} 
+ \hf \o_{a}{}^{ b c} (x) \,\cM_{ b c}~,
\label{WZ}
\ee
where $\nabla_{a}$ stands for  the covariant derivative
of anti-de Sitter space $\rm AdS_3$, 
\bea
{[} \nabla_{a},\nabla_{b} {]}&=&-4S^2 \cM_{ab}~.
\eea

We are interested in constructing 
an AdS  generalisation of the action (\ref{flat-Ac}). On general grounds, it 
should have  the form
\bea
S [ \cL^{(2)}] =S_0+\cdots~,~~~~~~
S_0=\frac{1}{8\p} \oint_{\g}  { v_i {\rm d} v^i }\int {\rm d}^3x\, e \,
\big(\cD^{(-2)}\big)^2 \big(\cD^{(0)}\big)^2 
\cL^{(2)}  ||
~,
\label{comp-Ac-000}
\eea
with $e:={\rm det}^{-1}{(e_m{}^a)}$. 
Note that in (\ref{comp-Ac-000})
the dots stand for curvature dependent corrections which are necessary for the action to be 
invariant under the symmetries of its parent action (\ref{InvarAc}).
It is interesting to note that there is one symmetry which is shared by 
the flat action (\ref{flat-Ac}) and the parent curved full superspace action (\ref{InvarAc}):
both are manifestly projective invariant (\ref{proj-inv}).
On the other hand $S_0$  is not projective invariant. 
As discussed in \cite{KT-M,KT-M_5D,KT-M-normal}, one can
actually exploit projective invariance
as a tool to iteratively find the completion of $S_0$ to $S[\cL^{(2)}]$ in  (\ref{comp-Ac-000}).
In Appendix A we sketch how to describe this approach for the (3,0) and (2,1) AdS cases.
Let us now write down the form of the full $\cN=3$ AdS projective action principle in
components
\bea
S[\cL^{(2)}]&=&
\frac{1}{8\p} \oint_{\g}  { v_i {\rm d} v^i }\int {\rm d}^3x\, e \,
\Big{[}
\big(\cD^{(-2)}\big)^2 \big(\cD^{(0)}\big)^2 
+4 \ri\big(\cS-2\cS^{(0)}\big)(\cD^{(-2)})^2
\non\\
&&~~~~~~~~~~~~~~~~~~
+12 \ri\cS^{(-2)}\cD^{(-2)\a}\cD_\a^{(0)}
-16\ri\cS^{(-4)}(\cD^{(0)})^2
\non\\
&&~~~~~~~~~~~~~~~~~~
-144\cS^{(-2)}\cS^{(-2)}
+64\cS^{(-4)}\cS^{(0)}
+48 \cS^{(-4)}\cS
\Big{]}\cL^{(2)}  ||
~.~~~~~~~~~
\label{components-Ac}
\eea
Here we have used the definitions
\bea
\cS^{(0)}:=\frac{v_iv_ju_ku_l \cS^{ijkl}}{(v,u)^2}~,~~~
\cS^{(-2)}:=\frac{v_iu_ju_ku_l \cS^{ijkl}}{(v,u)^3}~,~~~
\cS^{(-4)}:=\frac{u_iu_ju_ku_l \cS^{ijkl}}{(v,u)^4}~.
\eea
 The actions corresponding to the (3,0) and (2,1) AdS superspaces are obtained from 
(\ref{components-Ac}) by choosing the curvature as follows:
\bsubeq
\bea
(3,0):&&~
\cS=S~,~~
\cS^{ijkl}=0~,
\\
(2,1):&&~
\cS=\frac{1}{3}S~,~~
\cS^{ijkl}=-Sw^{(ij}w^{kl)}
~,~~
w^{ij}w_{ij}=2
~.
\eea
\esubeq

\section{Supersymmetric action: Reduction to $\cN=2$  superspace}
\label{N3toN2}
\setcounter{equation}{0}

The representation \eqref{components-Ac}  obtained in the previous subsection, 
corresponds to the situation when all the Grassmann integrals in the action (\ref{InvarAc})
have been done. Here we take a different course and reduce the superspace integral in   
(\ref{InvarAc})
to that over a certain $\cN=2$ subspace of the full $\cN=3$ AdS superspace under consideration. 
Such a procedure cannot be carried out in a unified way for  the cases (3,0) and (2,1), 
and thus a separate consideration should be given in each case.

\subsection{AdS superspace reduction: (3,0) to (2,0) }

To identify an $\cN=2$ subspace of the $\cN=3$ AdS superspace, 
we need a subset of four spinor covariant derivatives 
which, together with $\cD_a$, lead to  a closed set of (anti) commutation relations.

In the case of (3,0) AdS superspace, 
the covariant derivatives obey the (anti) commutation relations  \eqref{3_0-alg-isospinor}.
A  closed subalgebra  can be identified with 
the mutually conjugate derivatives $\cD_\a^{\1\1}$ and $-\cD_\a^{\2\2}$
(for any bosonic superfield $U$, it holds that  
$\overline{\cD_\a^{\1\1} U}=-\cD_\a^{\2\2} \overline{U}$).
Indeed, it follows from  \eqref{3_0-alg-isospinor} that 
\bsubeq \label{3_0-2_0}
\bea
&\{\cD_\a^{\1\1},\cD_\b^{\1\1}\}=
\{(-\cD_\a^{\2\2}),(-\cD_\b^{\2\2})\}=
0
~,~~~~~~
\label{3_0-2_0-1}
\\
&\{\cD_\a^{\1\1},(-\cD_\b^{\2\2})\}=
-2\ri\cD_{\a\b}
-4\ri\,\ve_{\a\b}\,S\,\cJ^{\1\2}
+4\ri\,S\,\cM_{\a\b}
~,~~~~~~
\label{3_0-2_0-2}
\\
&{[}\cD_{a},\cD_\b^{\1\1}{]}
=
\,S\,(\g_a)_\b{}^{\g}\cD_{\g}^{\1\1}
~,~~~
{[}\cD_{a},(-\cD_\b^{\2\2}){]}
=
\,S\,(\g_a)_\b{}^{\g}(-\cD_{\g}^{\2\2})
~,~~~
\label{3_0-2_0-3}
\\
&{[}\cD_a,\cD_b{]}=\,-\,4\,S^2\,\cM_{ab}~.
\label{3_0-2_0-4}
\eea
\esubeq
For the subset $(\cD_a, \cD_\a^{\1\1}, -\cD_\a^{\2\2})$ 
chosen, the original $R$-symmetry group SU(2) 
reduces to U(1), and the corresponding generator $\cJ^{\1\2} $ acts on the spinor derivatives as
\bea
[\cJ^{\1\2},\cD^{\1\1}_\a]=\cD^{\1\1}_\a
~,\qquad
[\cJ^{\1\2},(-\cD^{\2\2}_\a)]=-(-\cD^{\2\2}_\a)
~.
\eea 
The (anti) commutation relations \eqref{3_0-2_0} can be recognised as those corresponding to 
the (2,0) AdS superspace, $\rm AdS_{(3|2,0)}$,  studied in \cite{KT-M-2011}.

Now, we can embed the superspace $\rm AdS_{(3|2,0)}$ into $\rm AdS_{(3|3,0)}$.
Given a tensor superfield  $U(x,\q_{\imath \jmath})$ in  $\rm AdS_{(3|3,0)}$,
we define its projection
\bea
U|:=U(x,\q_{\imath \jmath})|_{\q_{\1\2}=0}~.
\label{N2red-1}
\eea
By definition,  $U|$  still depends on the Grassmann coordinates
$\q^\mu:=\q^\mu_{\1\1}$ and their complex conjugate $\qb^\mu=\q^\mu_{\2\2}$.
For the (3,0) AdS covariant derivatives 
\be
\cD_{{A}}=E_{{A}}{}^{{M}}\pa_{{M}}
+\hf\O_{{A}}{}^{bc}\cM_{bc}
+\hf\F_{{A}}{}^{kl}\cJ_{kl}~,
\label{N2red-2}
\ee
the projection is defined as
\bea
\cD_{{A}}|=E_{{A}}{}^{{M}}|\pa_{{M}}
+\hf\O_{{A}}{}^{bc}|\cM_{bc}
+\hf\F_{{A}}{}^{kl}|\cJ_{kl}
~.
\label{N2red-3}
\eea
Since the operators $\big(\cD_a,\,\cD_\a^{\1\1},\,-\cD_\a^{\2\2} \big)$ 
form a closed  algebra, 
which is isomorphic to that of the covariant derivatives for $\rm AdS_{(3|2,0)}$,
one can use the freedom to perform general coordinate, local Lorentz and SU(2) transformations
to chose a gauge in which
\bea
\cD_\a^{\1\1}|=\bfD_\a
~,~~~~~~
(-\cD_\a^{\2\2})|=\bfDB_\a
~,
\eea
where 
\bea
\bfD_A = (\bfD_a , \bfD_\a, \bar \bfD^\a)= {\bm E}_A{}^M \pa_M +\hf {\bm \O}_A{}^{cd}\cM_{cd}
+ \ri \,{\bm  \F}_A \cJ
\label{1.4}
\eea 
denote the covariant derivatives of $\rm AdS_{(3|2,0)}$ which obey the (anti) commutation 
relations \eqref{20AdSsuperspace}, with $\cJ\equiv \cJ^{\1\2}$.
In such a coordinate system\footnote{This is in fact a normal coordinate system for 
$\rm AdS_{(3|3,0)}$ around the submanifold $\rm AdS_{(3|2,0)}$.}  
the operators $\cD_\a^{\1\1}|$ and $\cD_{\a}^{\2\2}|$  involve no
partial derivative with respect to $\q_{\1\2}$, 
and therefore, for any positive integer $k$,  
it holds that $\big( \cD_{\hat{\a}_1} \cdots  \cD_{\hat{\a}_k} U \big)\big|
= \cD_{\hat{\a}_1}| \cdots  \cD_{\hat{\a}_k}| U|$, 
where $ \cD_{\hat{\a}} :=\big( \cD_\a^{\1\1}, -{\cD}_\a^{\2\2} \big)$ 
and $U$ is a tensor superfield. This implies that $\cD_a| = {\bf D}_a$.

Our next task is to reduce the transformation laws of projective supermultiplets 
from $\rm AdS_{(3|3,0)}$ to its $\cN=2$ subspace 
$\rm AdS_{(3|2,0)}$. Consider a Killing vector field of (3,0) AdS superspace, 
\bea
\x = \x^a \cD_a + \x^\a_{ij} \cD^{ij}_\a~.
\eea
We recall that $\x$ obeys the Killing equations \eqref{4.4}
which are equivalent to \eqref{2_1-Killing_iso_1-0} -- \eqref{2_1-Killing_iso_3-0}.
We introduce $\cN=2$ projections of the transformation parameters involved
\bsubeq
\bea
&&
\t^a:=\x^a|
~,~~~
\t^\a:=\x^\a_{\1\1}|~,~~~
\bar{\t}^\a=\x^\a_{\2\2}|~,~~~
t:=\ri\L^{\1\2}|=\overline{t}
~, ~~~t^{ab}:= \L^{ab}|~
;
\\
&&
\r^\a:=-\ri\x^\a_{\1\2}|
=\overline{\r^\a}
~,~~~~~~
\bar{\ve}:=\L^{\1\1}|~,~~~
\ve=\L^{\2\2}|=\L_{\1\1}|
~.
\eea
\esubeq

The important point is  that the parameters $(\t^a, \, \t^\a,\,\bar{\t}_\a,\, t^{\a\b} ,\, t )$
describe the infinitesimal isometries of  the (2,0) AdS superspace \cite{KT-M-2011}.
Such  transformations are  generated by the Killing vector fields,
$\t=\t^a\bfD_a+\t^\a\bfD_\a+\bar{\t}_\a\bfDB^\a$, obeying the Killing equation
\bea
\Big{[}\t+\ri t\cJ+\hf t^{bc}\cM_{bc},\bfD_A\Big{]}=0~,
\label{5.10K}
\eea
for some parameters $t$ and $t^{ab}$.
This equation  is equivalent to 
 \bsubeq \label{2_0-Killing_iso}
\bea
&& 4S \t_\a=\bfDB_\a t=\frac{2\ri}{3} S
\bfDB^\b\t_{\a\b}=\frac{\ri}{3}\bfDB^\b t_{\a\b}
~,
\label{2_0-Killing_iso_1}
\\
%%%
&&\bfDB_\a\t_\b=
\bfD_{(\a}\t_{\b\g)}=\bfD_{(\a} t_{\b\g)}
=0~,
\label{2_0-Killing_iso_2}
\\
 &&
\bfD_\g\t^\g
=
-\bfDB^\g\bar{\t}_\g
=2\ri t ~,
\label{2_0-Killing_iso_3}
\\
&&\bfD_{(\a}\t_{\b)}=-\bfDB_{(\a}\bar{\t}_{\b)}=\hf t_{\a\b}+ S \t_{\a\b} 
~.
\label{2_0-Killing_iso_4}
\eea
\esubeq
 These equations automatically follow from  the (3,0) Killing  equations,
eqs. \eqref{2_1-Killing_iso_1-0} -- \eqref{2_1-Killing_iso_3-0},  upon $\cN=2$ projection.
The real parameter $t |_{\q=0}=\rm const$ generates U$(1)_R$ transformations 
of  the (2,0) AdS superspace, where U$(1)_R$ is a subgroup of the $R$-symmetry group 
SU$(2)_R$ of  the (3,0) AdS superspace.

The transformation parameters $\r^\a$, $\ve $ and $\bar{\ve}$  generate
the third supersymmetry and those $R$-symmetry transformations which parametrise 
the coset SU$(2)_R / {\rm U}(1)_R$. Making use of \eqref{4.6}, one can show that 
$\r_\a$ is determined in terms of  $\ve$ and $\bar{\ve}$:
\bea
\r_\a=-\frac{1}{8S}\bfD_\a\ve=-\frac{1}{8S}\bfDB_\a\bar{\ve}
~.
\eea
The parameters $\ve$ and $\bar{\ve}$  satisfy the following properties
\bea
\bfD_\a\ve=\bfDB_\a\bar{\ve}~,~~~
\bfDB_\a\ve =0~, ~~~\bfD_{\a\b}\ve=0
~,~~~
\bfD^2\ve=-8\ri S\bar{\ve}
~.~~~~~~
\eea
These imply that the only independent components of $\ve$ are $\ve|_{\q=0}$ and 
$\bfD_\a \ve|_{\q=0}$.

The notion of $\cN=2$ projection is especially useful when dealing with projective multiplets. 
Given a covariant weight$-n$ projective multiplet $Q^{(n)}(v)$, 
it  can always be described  in terms of 
a related superfield $Q^{[n]} (\z)$ which depends on $\z$ and is proportional to the original 
superfield, $Q^{[n]} (\z) \propto Q^{(n)} (v)$. The precise definition of $Q^{[n]} (\z)$ depends 
upon the specific projective multiplet under consideration. 
Using $Q^{[n]}(\z)$, the analyticity constraint (\ref{ana}) becomes 
\bea
&&0=\z^2\cD_\a^{\1\1}Q^{[n]}(\z)
-2\z\cD_\a^{\1\2}Q^{[n]}(\z) 
+\cD_\a^{\2\2}Q^{[n]}(\z)
~,
\eea
or equivalently
\bea
\cD_\a^{\1\2}Q^{[n]}(\z) =\hf 
\Big( \z\cD_\a^{\1\1}
+\frac{1}{\z}\cD_\a^{\2\2}\Big)
Q^{[n]}(\z) 
~.
\label{an-north}
\eea
This equation shows that the dependence of $Q^{[n]} (x,\q_{\imath \jmath} , \z)$ on 
the Grassmann coordinates $\q^\m_{\1\2}$
 is completely determined in terms of its dependence on 
the other Grassmann coordinates  $\q^\m_{\1\1} $ and  $\q^\m_{\2\2} $. In other words, all 
information about the projective multiplet $Q^{[n]}( \z)$ 
is encoded in its $\cN=2$ projection $Q^{[n]}(\z)|$.

We now list the transformation laws of several projective multiplets under 
the (3,0) AdS  isometry group,  $\rm OSp (3|2; {\mathbb R} ) \times   Sp(2, {\mathbb R})$.
All multiplets will be projected to (2,0) AdS superspace, however will will not indicate explicitly the 
bar-projection.

We recall that a weight-$n$ projective superfield $Q^{(n)}$ transforms under the isometry group 
 $\rm OSp (3|2; {\mathbb R} ) \times   Sp(2, {\mathbb R})$ as
\bea
\d_\x Q^{(n)}(z,v)
&=& \Big(   \x^a\cD_a   +  \x_{kl}^\a \cD_\a^{kl}
- \hf\L^{(2)}{\bm \pa}^{(-2)}
+\frac{n}{2} \,\L^{(0)}
 \Big) Q^{(n)}(z,v) 
 ~,
\eea 
which follows from eq. (\ref{harmult1}).
Given an arctic weight-$n$ multiplet $\U^{(n)}(v)$, it can be conveniently  represented as 
\be
\U^{(n)}(v)=(v^{\1})^n\U^{[n]}(\z)~.
\label{5.17}
\ee 
Then $\U^{[n]}(\z)$ transforms as follows:
\bea
\d_\x \U^{[n]} 
&=& \Big{\{}
\, \t
+\ri t\Big( \,\z\frac{\pa}{\pa \z}
-\frac{n}{2} \Big)
+\ri\z\r^\a \bfD_\a
+\frac{\ri}{\z}\r_\a \bfDB^\a
+ \frac{1}{2}\Big(
 \frac{\ve}{\z} + \z\bar{\ve}
\Big)\z \frac{\pa}{\pa \z}
-\frac{n}{2} \,\z\bar{\ve}
 \Big{\}} \U^{[n]}
 ~.~~~~~~~~
 \label{Q-north}
\eea 
Given an antarctic weight-$n$ multiplet $\breve{\U}^{(n)}(v)$, it  is represented in the form 
\be
\breve{\U}^{(n)}(v)=(v^{\2})^n \breve{\U}^{[n]}(\z) =(v^{\1})^n \z^n\breve{\U}^{[n]}(\z)~.
\label{5.19}
\ee 
The transformation law of $\breve{\U}^{[n]}(\z)$ is 
\bea
\d_\x \breve{\U}^{[n]}
= \Big\{
\t
&+&\ri t\Big( \,\z\frac{\pa}{\pa \z}
+\frac{n}{2} \Big)
+\ri\z\r^\a \bfD_\a
+\frac{\ri}{\z}\r_\a \bfDB^\a
+ \frac{1}{2}\Big( \frac{\ve}{\z}
+\z \bar{\ve} 
 \Big) \z \frac{\pa}{\pa \z}
+\frac{n}{2\z} \, \ve
 \Big\} \breve{\U}^{[n]}
  ~.~~~~~~~~
\eea
Given a real weight-$(2n)$ multiplet ${G}^{(2n)}(v)$, $\breve{G}^{(2n)} = G^{(2n)}$, 
it is represented as
\be
{G}^{ (2n) }(v)=( \ri v^{\1}v^{\2} )^n {G}^{[2n]}(\z) =(v^{\1})^{2n} (\ri \,\z)^n  {G}^{[2n]}(\z)~.
\label{2.56}
\ee 
The transformation law of ${G}^{[2n]}(\z)$ is 
\bea
\d_\x {G}^{[2n]}
= \Big\{
\t
&+&\ri t  \,\z\frac{\pa}{\pa \z}
+\ri\z\r^\a \bfD_\a
+\frac{\ri}{\z}\r_\a \bfDB^\a
+ \frac{1}{2}\Big( \frac{\ve}{\z}
+\z \bar{\ve}  \Big) \z \frac{\pa}{\pa \z}
+\frac{n}{2} \Big( \frac{\ve}{\z} - \z \bar \ve\Big)
 \Big\} {G}^{[2n]}
  ~.~~~~~~~~
  \label{4.57}
\eea

To conclude the analysis of this subsection, we 
present the (3,0) supersymmetric  action reduced to (2,0) superspace.
In accordance with \eqref{2.56}, associated with the Lagrangian $\cL^{(2)}(v)$
is the superfield  $\cL^{[2]}(\z)$ defined 
by the rule $\cL^{(2)}(v) = \ri (v^{\1})^2 \z \cL^{[2]}(\z)$.
It turns out that the (3,0) supersymmetric action (\ref{InvarAc})
takes the following form in (2,0) AdS superspace
\bea
S[\cL^{(2)}] &=& \oint_\g  \frac{\rd \zeta}{2\pi \ri \zeta}
\int \rd^3x\, \rd^2{ \q} \rd^2{ \qb}\, {\bm E}\, 
 \cL^{[2]}
~,~~~~~~
{\bm E}^{-1}:={\rm Ber}({\bm E}_A{}^M)~.
\label{AdS-3_0-2_0_action}
\eea
To prove that  (\ref{AdS-3_0-2_0_action}) is the (2,0) reduction of the 
(3,0) action (\ref{InvarAc}) we check explicitly that it is invariant under the full isometry group 
of (3,0) AdS superspace, 
 $\rm OSp (3|2; {\mathbb R} )\times   Sp(2, {\mathbb R})$.
 Making  use of  (\ref{4.57}), the variation of (\ref{AdS-3_0-2_0_action}) is
\bea
\d_\x S[\cL^{(2)}] &=&\oint_\g  \frac{\rd \zeta}{2\pi \ri \zeta} 
 \int \rd^3x\, \rd^2{ \q} \rd^2{ \qb}\, {\bm E} 
 \Big{[} 
\t +\ri t \z\frac{\pa}{\pa \z} 
\non\\
&&
+\ri\z\r^\a \bfD_\a
+\frac{\ri}{\z}\r_\a \bfDB^\a
- \frac{1}{2}\z\bar{\ve} 
+ \frac{1}{2\z}\ve 
+ \big(\frac{1}{2}\z\bar{\ve}
+ \frac{1}{2\z}\ve \big)\z\frac{\pa}{\pa \z} 
 \Big{]} \cL^{[2]}
 ~.
\eea
The expression in the first line corresponds to the variation of $\cL^{[2]} $ under 
an infinitesimal  isometry transformation of (2,0) AdS superspace. Since the action is manifestly 
invariant under the (2,0) AdS isometry group, this variation vanishes.
For the remaining variation, upon integration by parts, we obtain
\bea
\d_\x S[\cL^{(2)}] &=& \oint_\g   \frac{\rd \zeta}{2\pi \ri } 
\int \rd^3x\, \rd^2{\q} \rd^2{ \qb}\, {\bm E} 
 \Big( 
\big(
\ri ( \bfD_\a\r^\a)
- \bar{\ve}\big)
+\frac{1}{\z^2}\big(
\ri(\bfDB^\a\r_\a)
+\ve \big)
 \Big) \cL^{[2]}
=0~,~~~~~~~~~
\eea
which is identically zero due to the identities
\bea
\ri\bfD_\a\r_\b=-\hf\ve_{\a\b}\,\bar{\ve}~,~~~~~~
\ri\bfDB_\a\r_\b=-\hf\ve_{\a\b}\,\ve
~.
\label{3_0-extra}
\eea
We conclude by noticing that the auxiliary superfield $\cC^{(-4)}$  (\ref{InvarAc}) 
has dropped out upon reduction to (2,0) AdS  superspace.

%%%%%%%%%%%%%%%%%%%%%%%%%%%%%%%%%%%%%%%%

\subsection{AdS superspace reduction: (2,1) to (2,0)}

We now turn to developing $\cN=2$ reduction schemes for the projective multiplets in (2,1) AdS 
superspace. 
The main difference between the (3,0) and (2,1) AdS superspaces is that the latter possesses 
the  covariantly constant  tensor $w^{ij}$ (which can be interpreted as the field strength of a frozen
$\cN=3$ vector multiplet). As follows from the algebra of (2,1) AdS covariant derivatives, 
eq. \eqref{2_1-alg-isospinor}, 
the $R$-symmetry part of the holonomy  group of this superspace
is no longer  SU$(2)_R$, as in the (3,0) case; instead it is the group U$(1)_R$ which 
is associated with the generator by $ \cJ=- \frac{\ri}{2} w^{ij}\cJ_{ij}$. 
Therefore, the local SU$(2)_R$ group can be used to choose the SU$(2)_R$ connection 
to be  
\bea
\F_A^{kl}=w^{kl}\F_A~.
\label{5.27}
\eea 
In this gauge the tensor $w^{ij}$ becomes strictly constant, $w^{ij}=\rm const$, and turn into
an invariant tensor of the (2,1) AdS isometry group 
$\rm OSp (2|2; {\mathbb R} ) \times   OSp(1|2, {\mathbb R})$.
It turns out that different numeric choices for $w^{ij}$ 
correspond to the possibility to perform reduction either  
to the (2,0) AdS superspace or to the (1,1) one. 
In other words, the (2,1) AdS superspace allows two inequivalent $\cN=2$ reduction schemes. 

Here we focus on the AdS reduction $ (2,1) \to (2,0)$. If we choose
\bea
w^{\1 \1}=w^{\2 \2}=0~,~~~
w^{\1\2}=-w_{\1 \2}=-\ri
\label{3_0-2_0-start}
\eea
in the (2,1) algebra (\ref{2_1-alg-isospinor}), then the operators
$\cD_\a^{\1\1}$ and $(-\cD_\a^{\2\2})$
can be seen to satisfy the same (anti) commutation relations  as 
eqs. (\ref{3_0-2_0-1})--(\ref{3_0-2_0-3}) which are 
equivalent to the (2,0) AdS algebra (\ref{AdS_(2,0)_algebra_1})--(\ref{AdS_(2,0)_algebra_2}).
Therefore, the projection from (2,1) to (2,0) AdS formally proceeds exactly as in the (3,0), 
see the analysis around the equations (\ref{N2red-1})--(\ref{N2red-3}).
The only difference is that in (\ref{N2red-1}) and (\ref{N2red-2})
the (2,1) connections should be as in \eqref{5.27}.

Consider the Killing vector fields, $\x^A = (\x^a , \x^\a_{ij})$,  of the (2,1) AdS superspace.  
They obey the Killing equations, eq.  \eqref{4.4}, and hence  
\bsubeq
\bea
\cD_\a^{ij}\x_\b^{kl}
&=&
\hf\ve_{\a\b}\big(\ve^{ik}w^{jl}
+\ve^{jl}w^{ik}\big)\L
-S(\ve^{i(k}\ve^{l)j}+w^{ij}w^{kl})\x_{\a\b}
-\hf\ve^{i(k}\ve^{l)j}\L_{\a\b}
~,
\label{2_1-Killing_iso_1}
\\
%%%%%%
0&=&
\cD_{ \g}^{ij}\x^{\a\g}
+6\ri\x^{\a ij}
~,~~~
0=
\cD_\g^{ij}\L^{\a\g}
-12\ri S\x^{\a}_{kl}(\ve^{i(k}\ve^{l)j}+w^{ij}w^{kl})
\label{2_1-Killing_iso_2}~,
\\
%%%%%%
0&=&
\cD^{ij}_{(\a }\x_{\b\g)}
=\cD^{ ij}_{ (\a }\L_{\b\g)}
~,
\label{2_1-Killing_iso_3}
\eea
\esubeq
where we have used the fact that  
the $R$-symmetry group in the AdS (2,1) case  reduces to U$(1)_R$
and the corresponding transformation parameter is 
\bea
\L^{ij}=w^{ij}\L
~,~~~
\L=\hf w_{kl}\L^{kl}
~.
\eea
We now project  the transformation parameters to (2,0) superspace
\begin{subequations}
\bea
&&\t^a:=\x^a|
~,~~~
\t^\a:=\x^\a_{\1\1}|~,~~~
\bar{\t}^\a:=\x^\a_{\2\2}|~,~~~
t:=\ri\L^{\1\2}|=\L|=\overline{t}
~, ~~~ t^{ab} := \L^{ab}|
~;~~~~~~
\\
&&\r^\a:=-\ri\x^\a_{\1\2}|
=\overline{\r^\a}
~.
\label{Kill-red-3}
\eea
\end{subequations}
Because of (\ref{3_0-2_0-start}), it holds that
$\L^{\1\1}=\L^{\2\2}=0$, which is clearly different from the (3,0) case.
As in the (3,0) case, the parameters $(\t^a, \, \t^\a,\,\bar{\t}_\a,\, t^{\a\b} ,\, t )$
describe the infinitesimal isometries of  the (2,0) AdS superspace.
We recall that such  transformations are  generated by the Killing vector fields,
$$
\t=\t^a\bfD_a+\t^\a\bfD_\a+\bar{\t}_\a\bfDB^\a~,
$$ 
obeying the equations \eqref{5.10K} or, equivalently, 
\eqref{2_0-Killing_iso}. 
The real spinor parameter $\r^\a$ generates
the third supersymmetry  transformation.
Making use of eqs.   (\ref{2_1-Killing_iso_1})--(\ref{2_1-Killing_iso_3}) gives
\bea
\bfD_\a\r_\b = \bfDB_\a\r_\b=0~.
\label{2_1-extra}
\eea
These conditions mean that $\r_\a$ is an ordinary  Killing spinor, 
\bea
\bfD_{\b\g} \r_\a = S(\ve_{\a \b } \r_\g + \ve_{\a\g} \r_\b)~.
\eea

To complete the AdS superspace reduction $(2,1) \to (2,0)$, 
it remains to work out the transformation laws of projective multiplets
under the (2,1) AdS isometry group,  $\rm OSp (2|2; {\mathbb R} ) \times   OSp(1|2, {\mathbb R})$.
In (2,1) AdS superspace,  
a covariant weight-$n$ projective multiplet $Q^{(n)}$ transforms as 
\bea
\d_\x Q^{(n)}(z,v)
&=& \Big( \x^a\cD_a   + \x_{kl}^\a \cD_\a^{kl}
- \hf w^{(2)}\L{\bm \pa}^{(-2)}
+\frac{n}{2} \,w^{(0)}\L
 \Big) Q^{(n)}(z,v) 
 ~,
 \label{Q-iso-2_1}
\eea 
in accordance with  eq. (\ref{harmult1}).
We project this transformation law to (2,0) AdS superspace.
Given an arctic  weight-$n$ multiplet $\U^{(n)} (v)$, we associated with it 
the superfield $\U^{[n]}(\z)$ defined by \eqref{5.17}. The latter transforms as follows:
\bea
\d_\x \U^{[n]}
&=& \Big{\{}\,\t
+\ri t \Big(\z\frac{\pa}{\pa \z}
-\frac{n}{2} \Big)
+\ri\z\r^\a \bfD_\a
+\frac{\ri}{\z}\r_\a \bfDB^\a
 \Big{\}} \U^{[n]}
 ~.~~~~~~~~~
\label{5.35}
\eea 
Given an antarctic  weight-$n$ multiplet $\breve{\U}^{(n)} (v)$, we associated with it 
the superfield $\breve{\U}^{[n]}(\z)$ defined by \eqref{5.19}. The latter transforms as follows:
\bea
\d_\x  \breve{\U}^{[n]}
&=& \Big{\{}\, \t
+\ri t \Big(\z\frac{\pa}{\pa \z}
+\frac{n}{2} \Big)
+\ri\z\r^\a \bfD_\a
+\frac{\ri}{\z}\r_\a \bfDB^\a
 \Big{\}} \breve{\U}^{[n]}
 ~.~~~~~~~~~
\eea 
Given a real weight-$(2n)$ multiplet ${G}^{(2n)}(v)$, $\breve{G}^{(2n)} = G^{(2n)}$, 
we associate with it the superfield ${G}^{[2n]}(\z)$, eq. \eqref{2.56},  
with the transformation law
\bea
\d_\x G^{[2n]}
&=& \Big{\{}\,\t
+\ri t \ \z\frac{\pa}{\pa \z}
+\ri\z\r^\a \bfD_\a
+\frac{\ri}{\z}\r_\a \bfDB^\a
 \Big{\}} G^{[2n]}
 ~.~~~~~~~~~
\eea

To conclude the subsection we note that the  (2,1) AdS supersymmetric action 
reduced to (2,0) AdS superspace
 has exactly the same form as the (3,0) case: eq. (\ref{AdS-3_0-2_0_action}).
 The proof that the action of the form (\ref{AdS-3_0-2_0_action}) is invariant
 under the (2,1) isometries reduced to (2,0), up to minor differences,
 goes along the same line of the (3,0) case.

%%%%%%%%%%%%%%%%%%%%%%%%%%%%%%%%%%%%%%%%

\subsection{AdS superspace reduction: (2,1) to (1,1)}

AdS superspace reduction $(2,1) \to (1,1)$ corresponds to the following choice of $w^{ij}$:
\bea
&&w^{\1\2}=0~,~~~
w:=w^{\1\1}~,~~~
\bar{w}=w^{\2\2}=w_{\1\1}~,~~~
|w|^2=1
~.
\label{5.38}
\eea
Making use of this $w^{ij}$ in the (anti) commutation relations 
(\ref{2_1-alg-isospinor-1})--(\ref{2_1-alg-isospinor-3}),
and also introducing new AdS parameters
\bea
\m= \ri S\bar{w}^2~,~~~
\mub=-\ri Sw^2~,
\eea
we get the algebra
\bsubeq
\bea
&\{\cD_\a^{\1\1},\cD_\b^{\1\1}\}=
-4\mub\cM_{\a\b}
~,~~~~~~
\{(-\cD_\a^{\2\2}),(-\cD_\b^{\2\2})\}=
4\mu\cM_{\a\b}
~,
\label{alg-2_1-to-1_1a}
\\
%%%%%%%%%%%%
&\{\cD_\a^{\1\1},(-\cD_\b^{\2\2})\}=
-2\ri\cD_{\a\b}
~,
~~~~~~
{[}\cD_{a},\cD_\b^{\1\1}{]}
=
\ri\mub\,(\g_a)_\b{}^\g(-\cD_{\g}^{\2\2})
~,
\label{alg-2_1-to-1_1b}
%%%%%%
\\
&{[}\cD_{a},(-\cD_\b^{\2\2}){]}
=
-\ri\mu\,(\g_a)_\b{}^\g\cD_{\g}^{\1\1}
~,~~~~~~
{[}\cD_{a},\cD_b{]}
=
-4\,|\mu|^2\,\cM_{ab}
~.
\label{alg-2_1-to-1_1c}
\eea
\esubeq
The (anti) commutation relations coincide with those corresponding to the covariant derivatives
of (1,1) AdS superspace, 
eqs. (\ref{1_1-alg-AdS-2-1b})--(\ref{1_1-alg-AdS-2}).
Since no U$(1)_R$ curvature is present in the relations 
\eqref{alg-2_1-to-1_1a}--\eqref{alg-2_1-to-1_1c}, 
we can use the local U$(1)_R$ symmetry to choose a gauge in which the covariant derivatives
$\cD_a, \cD_\a^{\1\1} $ and $\cD_\a^{\2\2}$ have no U$(1)_R$ connection. 

The AdS superspace projection $(2,1) \to (1,1)$  formally proceeds exactly as in the (3,0), 
eqs. (\ref{N2red-1})--(\ref{N2red-3}) with few differences:\\
${\,}\quad$(i) the  connection $\F_A{}^{kl}$ in (\ref{N2red-2})
 should be as in \eqref{5.27};\\
 ${}\quad$(ii) the general  coordinate invariance can be used to choose a gauge 
 \bea
 \cD_\a^{\1\1}|:=\de_\a~,~~~~~~
-  \cD_\a^{\2\2}|:=\deb_\a
~,
 \eea
where 
\bea
\de_A = (\de_a , \de_\a ,\deb^\a) =\cE_A{}^M \pa_M + \hf \O_A{}^{cd}\cM_{cd} 
\label{5.42}
\eea
are the covariant derivatives for (1,1) anti-de Sitter superspace, which obey the (anti) commutation 
relations (\ref{1_1-alg-AdS-2-1b})--(\ref{1_1-alg-AdS-2}).

Consider the Killing vector fields, $\x^A = (\x^a , \x^\a_{ij})$, 
 of the (2,1) AdS superspace.  
They obey the Killing equations \eqref{4.4} in which $\L^{ij}$ should be chosen in the form 
$\L^{ij} = w^{ij} \L$ with $w^{ij }$ given by eq. \eqref{5.38}.
We project the transformation parameters to (1,1) AdS superspace:
\bsubeq
\bea
&
l^a=\x^a|~,~~~
l^\a:=\x^\a_{\1\1}|~,~~~
\bar{l}^\a=\x^\a_{\2\2}|~,~~~\l^{ab}:= \L^{ab}|~;
~~~~~~
\\
&\r^\a:=-\ri\x^\a_{\1\2}|
=\overline{\r^\a}
~,~~~~~~
\ve:=\L|=\bar{\ve}
~.
\eea
\esubeq
The superfields $(l^a,\, l^\a,\,\bar{l}_\a,\, \l^{ab})$
describe an infinitesimal isometry transformation of the (1,1) AdS superspace \cite{KT-M-2011}.
 The isometries are generated by (1,1) AdS Killing vector fields, 
 \bea
l =l^a\de_a+l^\a\de_\a+{\bar l}_\a\deb^\a~,
 \eea 
which are defined to obey the equations
\bea
\Big{[}l+\hf\l^{ab}\cM_{ab},\de_C\Big{]}=0~,
\eea
which are equivalent to 
\begin{subequations}\label{1,1-SK}
\bea
0&=&
\de_{(\a}l_{\b)}
-\hf\l_{\a\b}
~,~~~
0=
\deb_{(\a}l_{\b)}
+\ri\mu\, l_{\a\b}
~,~~~
\de_\a l^\a=\deb^\a l_\a=0~,
\label{1,1-SK_1}
\\
%%%%%%
0&=&
\de^\b\l_{\a\b}
-12\mub\, l_{\a}~,
~~~
0=
\deb^\b l_{\a\b}
+6\ri\, l_{\a}~,
~~~
\de_{(\a} l_{\b\g)}=\de_{(\a}\l_{\b\g)}=0
~.
\label{1,1-SK_2}
\eea
\end{subequations}
The (1,1) AdS Killing vector fields can be shown to generate the supergroup 
$\rm OSp(1|2;{\mathbb R}) \times OSp(1|2;{\mathbb R})$.
The relations \eqref{1,1-SK} follow
 by projecting the (2,1) Killing vector equations,
 (\ref{2_1-Killing_iso_1})--(\ref{2_1-Killing_iso_3}), to the (1,1) AdS superspace. 

The  parameters $\r^\a|_{\q=0} $ and $\ve|_{\q=0}$  generate
the third supersymmetry and  U(1) transformations respectively.
By using  (\ref{2_1-Killing_iso_1})--(\ref{2_1-Killing_iso_3}), one can derive the following
equations
\bea
\ri\de_\a\r_\b
=
-\hf\ve_{\a\b}\,w\,\ve
~,~~~~~~
\ri\deb_\a\r_\b
=
-\hf\ve_{\a\b}\,\bar{w}\,\ve
~.
\label{2_1-extra-2}
\eea
It can be further shown that the spinor superfield $\r_\a$ is determined in terms of $\ve$ as
\bea
\r_\a =-\frac{1}{4S w}\de_\a\ve=-\frac{1}{4S\bar{w}}\deb_\a\ve
~,
\eea
where $\ve$ can be proven to satisfy the equations
\bsubeq
\bea
&
\bar{w}\de_\a\ve=w\deb_\a\ve~,~~~
(\de^2-4\mub ) \ve=0~,~~~
(\deb^2 - 4\mu )\ve=0~,
\\
&(\ri  \de^\a\deb_\a -4 |\m| )\ve =0
~,~~~
\de_{(\a}\deb_{\b)}\ve=\de_{\a\b}\ve=0~.
\eea
\esubeq

To complete the AdS superspace reduction $(2,1) \to (2,0)$, 
it remains to work out the transformation laws of projective multiplets
under the (2,1) AdS isometry group,  $\rm OSp (2|2; {\mathbb R} ) \times   OSp(1|2, {\mathbb R})$.
In (2,1) AdS superspace,  
a covariant weight-$n$ projective multiplet $Q^{(n)}$ transforms as in \eqref{Q-iso-2_1}.
We project the transformation law \eqref{Q-iso-2_1} to the (1,1) AdS superspace. 
Given an arctic  weight-$n$ multiplet $\U^{(n)} (v)$, we associated with it 
the superfield $\U^{[n]}(\z)$ defined by \eqref{5.17}. The latter transforms as follows:
\bea
\d_\x \U^{[n]}
&=& \Big{\{} l
+\ri\z\r^\a \de_\a
+\frac{\ri}{\z}\r_\a \deb^\a
+\hf  \ve \Big( w \z  + \frac{\bar{w}}{\z} \Big)  \z \frac{\pa}{\pa \z}
-\frac{n}{2}\ve \, w\z 
 \Big{\}} \U^{[n]}
 ~.~~~~~~
 \label{Q-north-3}
\eea 
Given an antarctic  weight-$n$ multiplet $\breve{\U}^{(n)} (v)$, we associated with it 
the superfield $\breve{\U}^{[n]}(\z)$ defined by \eqref{5.19}. The latter transforms as follows:
\bea
\d_\x  \breve{\U}^{[n]}
&=& \Big{\{} l
+\ri\z\r^\a \de_\a
+\frac{\ri}{\z}\r_\a \deb^\a
+\hf  \ve \Big( w \z  + \frac{\bar{w}}{\z} \Big)  \z \frac{\pa}{\pa \z}
+\frac{n}{2} \ve\, \frac{\bar w}{\z} 
 \Big{\}}  \breve{\U}^{[n]}
 ~.~~~~~~
\eea 
Given a real weight-$(2n)$ multiplet ${G}^{(2n)}(v)$, $\breve{G}^{(2n)} = G^{(2n)}$, 
we associate with it the superfield ${G}^{[2n]}(\z)$, eq. \eqref{2.56},   
with the transformation law
\bea
\d_\x G^{[2n]}
&=& \Big{\{} l
+\ri\z\r^\a \de_\a
+\frac{\ri}{\z}\r_\a \deb^\a
+\hf  \ve \Big( w \z  + \frac{\bar{w}}{\z} \Big)  \z \frac{\pa}{\pa \z}
+\frac{n}{2} \ve \Big( \frac{\bar w}{\z} -w \z  \Big) 
 \Big{\}} G^{[2n]}
 ~.~~~~~~
 \label{5.52}
\eea

Now let us show that  the (1,1) AdS supersymmetric action in (1,1) AdS superspace takes the form 
\bea
S[\cL^{(2)}] &=&\oint_\g   \frac{\rd \zeta}{2\pi \ri \zeta}
 \int \rd^3x\, \rd^2\q \rd^2{\qb}\, {\cE}\, 
 \cL^{[2]}
~,~~~~~~
{\cE}^{-1}:={\rm Ber}({\cE}_A{}^M)~,
\label{AdS-2_1-1_1_action}
\eea
where $(x,{\q}^\mu,{\qb}_\mu)$ are the local coordinates on (1,1) AdS superspace, and  
${\cE}_A{}^M$ 
is the vielbein, eq. \eqref{5.42}. 
The Lagrangian $\cL^{[2]}$ 
is defined as usual, $\cL^{ (2) }(v) = \ri (v^\1)^2\z \cL^{[2]}(\z)$.
We have to demonstrate that the action is invariant under 
 the (2,1) AdS isometry group $\rm OSp (2|2; {\mathbb R} ) \times   OSp(1|2, {\mathbb R})$.
 By using the transformation law (\ref{5.52}) for $n=1$, the variation of 
 (\ref{AdS-2_1-1_1_action}) is
\bea
\d_\x S[\cL^{[2]}] &=&\oint_\g   \frac{\rd \zeta}{2\pi \ri \zeta}
 \int \rd^3x\, \rd^2{\q} \rd^2{\qb}\, {\cE}\, 
  \Bigg{[} 
l
\non\\
&&
+\ri\z\r^\a \de_\a
+\frac{\ri}{\z}\r_\a \deb^\a
- \frac{w}{2}\z \ve 
+ \frac{\bar{w}}{2\z} \ve
+\hf \Big(\z w 
+ \frac{\bar{w}}{\z} \Big)\ve\z\frac{\pa}{\pa \z} 
 \Bigg{]} \cL^{[2]}
 ~.
\eea
The variation in the first line does not contribute to $\d_\x S[\cL^{[2]}]$, since the action is 
manifestly (1,1) AdS supersymmetric.  
Integrating by parts in the second line gives
\bea
\d_\x S[\cL^{[2]}] &=&
\oint_\g   \frac{\rd \zeta}{2\pi \ri }
 \int \rd^3x\, \rd^2{\q} \rd^2{\qb}\, {\cE}\, 
  \Big( 
\big(
\ri(\de_\a\r^\a)
-w \ve\big)
+\frac{1}{\z^2}\big(
\ri(\deb^\a\r_\a)
+\bar{w}\ve \big)
 \Big) \cL^{[2]}
=0~,~~~~~~~~~
\eea
which is identically zero due to (\ref{2_1-extra-2}).

%%%%%%%%%%%%%%%%%%%%%%%%%%%%%%%%%%%%%%%%%%%%%%%%

\section{$\cN=3$ supersymmetric  sigma models in AdS}\label{N3-compo}
\setcounter{equation}{0}

We are now prepared to apply the formalism developed above to construct general (3,0) and (2,1) 
supersymmetric
nonlinear $\s$-models in $\rm AdS_3$.

\subsection{Sigma models with (3,0) AdS supersymmetry}

By analogy with the rigid supersymmetric case, it is natural to expect that a general nonlinear 
$\s$-model
with (3,0) AdS supersymmetry can be realised in terms of covariant weight-one arctic multiplets
$\U^{(1) \,I} (v) $ and their smile-conjugates 
$\breve{\U}^{(1)\,\bar I} (v)$,
with $I=1,\dots, n$.  
What can be said about the Lagrangian $\cL^{(2)}$ of such a theory? The specific feature of the 
(3,0) AdS
superspace is that there are no {\it background} projective multiplets which are invariant under 
the isometry supergroup  $\rm OSp (3|2; {\mathbb R} ) \times   Sp(2, {\mathbb R})$.\footnote{The 
situation is 
completely different in the (2,1) AdS case where the background $\cO(2)$ multiplet 
$w^{(2)} :=  v_i v_j w^{ij}$ 
is invariant under the isometry group 
$\rm OSp (2|2; {\mathbb R} ) \times   OSp(1|2; {\mathbb R})$.}
In order for $\cL^{(2)}$ to be a covariant weight-two projective multiplet, it cannot depend explicitly 
on $v^i$. 
It must be a function of the dynamical superfields only, 
\bea
\cL^{(2)} 
= {\rm i} \, K (\U^{(1)}, \breve{\U}^{(1)})~,
\label{sigma-3_0}
\eea
where  $K(\F^I, {\bar \F}^{\bar J}) $ is a homogeneous function of its arguments of degree one, 
\bea
\Big( \F^I \frac{\pa}{\pa \F^I} +  \bar \F^{\bar I} \frac{\pa}{\pa\bar  \F^{\bar I}} \Big)
K(\F, \bar \F) =  2K( \F,   \bar \F)~.
\label{5.2}
\eea
In order for  $\cL^{(2)}$ to be real with respect to the smile-conjugation
$$
\breve{}: ~~\U^{(1)} 
\to
\breve{\U}^{(1)}~, \qquad  \breve{\U}^{(1)} 
\to
- {\U}^{(1)}~,
$$
it suffices to subject $ K( \F,   \bar \F)$ to  additional conditions  \cite{K-comp,K-conf}
\bea
 \F^I \frac{\pa}{\pa \F^I} 
K(\F, \bar \F) =  K( \F,   \bar \F), \qquad \overline{K}=K~.
\label{Kkahler22}
\eea
This condition means that $K(\F, \bar \F) $ can be interpreted as the K\"ahler potential 
of a K\"ahler cone, see e.g.   \cite{GR}. By definition, this is a K\"ahler manifold $(\cM, g_{I\bar J})$
possessing a homothetic conformal Killing vector $\c$
 \bea
\c = \c^I \frac{\pa}{\pa \F^I} + {\bar \c}^{\bar I}  \frac{\pa}{\pa {\bar \F}^{\bar I}}
\equiv \c^\m \frac{\pa}{\pa \vf^\m} ~,
\eea
with the property
\bea
\nabla_\n \c^\m = \d_\n{}^\m \quad \Longleftrightarrow \quad 
\nabla_J \c^I = \d_J{}^I~, \qquad 
\nabla_{\bar J} \c^I = \pa_{\bar J} \c^I = 0~.
\label{hcKv}
\eea
In particular,  $\c $ is holomorphic. Its properties include:
\bea
{ g}_{I \bar J} \, \c^I {\bar \c}^{\bar J} ={ K}~, 
\qquad \c_I := {g}_{I \bar J} \,{\bar \c}^{\bar J} = \pa_I { K}\quad \Longrightarrow \quad 
\c^IK_I = K~, 
\label{hcKv-pot}
\eea
with $K$ the K\"ahler potential. Local complex coordinates for $\cM$ can always be
 chosen such that 
 \bea
\c = \F^I \frac{\pa}{\pa \F^I} + {\bar \F}^{\bar I}  \frac{\pa}{\pa {\bar \F}^{\bar I}}~, 
\eea
which correspond to our specific case, eq. \eqref{Kkahler22}.

In 3D $\cN=3$ flat projective superspace, any nonlinear $\s$-model with Lagrangian 
specified by eqs. \eqref{sigma-3_0} and 
 \eqref{Kkahler22} is $\cN=3$ superconformal \cite{KPT-MvU-2011} (which is a generalisation 
 of the earlier results in the 4D $\cN=2$ case  \cite{K-comp,K-conf}). 
 The target spaces of these $\s$-models are hyperK\"ahler cones, see e.g. \cite{GR,deWRV} and 
 references therein. 
 Since (3,0) AdS superspace is conformally related 
 to $\cN=3$ Minkowski superspace, 
we conclude that general nonlinear $\s$-models in (3,0) AdS superspace are $\cN=3$ 
superconformal. 

Consider the $\s$-model 
\bea
S &=& \oint_\g  \frac{\rd \zeta}{2\pi \ri \zeta}
\int \rd^3x\, \rd^2{ \q} \rd^2{ \qb}\, {\bm E}\, 
\cL^{[2]}
~, 
\eea
where 
\bea
\cL^{[2]} 
:= 
\frac{1}{\z}  K (\U^{[1]} , \z \breve{\U}^{[1]} ) 
\label{5.9}
\eea
At the moment we assume only the homogeneity condition \eqref{5.2}.
The transformation law of $\cL^{[2]} $ must be
\bea
\d_\x \cL^{[2]} 
= \Big\{
\t
+\ri t \,\z\frac{\pa}{\pa \z}
+ \ri\z\r^\a \bfD_\a
+\frac{\ri}{\z}\r_\a \bfDB^\a
+ \frac{1}{2}\Big(\z \bar{\ve} + \frac{1}{\z} \ve\Big) \z \frac{\pa}{\pa \z}
-\frac{1}{2} \,\z \bar\ve +\frac{1}{2\z} \, \ve
 \Big\} \cL^{[2]}~.~~~~~~~~~
 \label{5.10}
\eea
This should be induced by the variations of $\U^{[1]}$ and $\breve{\U}^{[1]} $ in \eqref{5.9}, 
which are 
\begin{subequations}
\bea
\d_\x \U^{[1]}
= \Big\{
\t
&+&\ri t\Big( \,\z\frac{\pa}{\pa \z}
-\frac{1}{2} \Big)
+ \ri\z\r^\a \bfD_\a
+\frac{\ri}{\z}\r_\a \bfDB^\a
+ \frac{1}{2}\Big(\z \bar{\ve}+ \frac{1}{\z} \ve\Big) \z \frac{\pa}{\pa \z}
-\frac{1}{2} \,\z\bar{\ve}
 \Big\} \U^{[1]}
 ~,~~~~~~~~~\\
\d_\x \breve{\U}^{[1]}
= \Big\{
\t
&+&\ri t\Big( \,\z\frac{\pa}{\pa \z}
+\frac{1}{2} \Big)
+\ri\z\r^\a \bfD_\a
+\frac{\ri}{\z}\r_\a \bfDB^\a
+ \frac{1}{2}\Big(\z \bar{\ve} + \frac{1}{\z} \ve\Big) \z \frac{\pa}{\pa \z}
+\frac{1}{2\z} \, \ve
 \Big\} \breve{\U}^{[1]}
~.~~~~~~~~~
\eea 
\end{subequations}
It is a short calculation to show that $\cL^{[2]} $ given by eq. \eqref{5.9}
transforms as in \eqref{5.10} if eq. \eqref{5.2} holds. 
On the other hand, the Lagrangian \eqref{5.9} is real under the smile 
conjugation
provided the stronger conditions \eqref{Kkahler22} hold.

\subsection{Sigma models with (2,1) AdS supersymmetry}

Unlike the (3,0) AdS superspace studied above, 
the (2,1) AdS superspace 
possesses a nontrivial covariantly constant tensor -- 
the  $\cO(2)$ multiplet $w^{(2)}=v_iv_jw^{ij}$, 
with $w^{ij}$ the parameter of the (2,1) AdS algebra
(\ref{2_1-alg-isospinor-1})--(\ref{2_1-alg-isospinor-3}). 
This invariant tensor can be used to construct supersymmetric theories
generated by Lagrangians  of the form\footnote{Similar models 
exist in 5D $\cN=1$ AdS \cite{KT-M} and 4D $\cN=2$ AdS \cite{KT-M_4D-conf-flat}.}
\bea
\cL^{(2)}=w^{(2)}\cL^{(0)}
~,
\eea
for some covariant real weight-zero projective multiplet $\cL^{(0)}$.

In (2,1) AdS superspace, 
general nonlinear $\s$-models can be described in terms of covariant weight-zero 
arctic  multiplets $\U^{ I}(v) $ and their smile-conjugates $\breve{\U}^{\bar I}(v)$ 
using the Lagrangian
\bea
\cL^{(2)}= w^{(2)}K (\U^{ I}, \breve{\U}^{\bar J}) ~,
\label{sigma-2_1}
\eea
where $K (\F^{ I}, \bar{\F}^{\bar J})$  is the K\"ahler potential of a real analytic K\"ahler manifold 
$\cX$.
The interpretation of $K$ as a K\"ahler potential is consistent, since the action 
generated by \eqref{sigma-2_1} turns out to be invariant under K\"ahler transformations of the form
\be
{ K}({ \U}, \breve{\U})~\to ~{ K}({ \U}, \breve{ \U})
+{ \L}({\U}) +{\bar { \L}} (\breve{\U} )~,
\label{1.26}
\ee
with ${ \L}(\F^I)$ a holomorphic function.
The target space $\cM$ of this $\s$-model proves to be an open domain of the zero section 
of the cotangent bundle of $\cX$, $\cM \subset T^*\cX$. This can be shown by generalizing 
the flat-superspace considerations of \cite{K98,GK2}. 

In general, $K (\F , \bar{\F} )$ in \eqref{sigma-2_1}
is an arbitrary real analytic function of $n$ complex variables. In the case that  
 $K (\F , \bar{\F} )$ obeys the homogeneity condition \eqref{Kkahler22}, 
 the Lagrangian \eqref{sigma-2_1} proves to define an $\cN=3$ superconformal 
 $\s$-model. Such a theory can be re-formulated entirely in terms of covariant weight-one 
 arctic multiplets, $\U^{(1)\, I} (v)$,  and their smile-conjugates in complete analogy with the 
 four-dimensional $\cN=2$ supersymmetric $\s$-models in AdS \cite{BKLT-M}.  
This requires to make use of an intrinsic hypermultiplet, $q^i$,  associated with the 
(2,1) AdS superspace. This hypermultiplet is defined in complete analogy 
with the 4D consideration given in section 2.2 of \cite{BKLT-M}.

It has been shown in the previous section that the (2,1) AdS superspace
allows two types of $\cN=2$ reduction, depending on the choice of $w^{ij}$ made. 
Any field theory in AdS${}_{(3|2,1)}$
can be reformulated as a dynamical system in AdS${}_{(3|2,0)}$ or in AdS${}_{(3|1,1)}$.
Upon reduction  to the (2,0) AdS superspace, 
the  supersymmetric $\s$-model \eqref{sigma-2_1} proves to be described by the 
action
\bea
S &=& \oint_\g  \frac{\rd \zeta}{2\pi \ri \zeta}
\int \rd^3x\, \rd^2{ \q} \rd^2{ \qb}\, {\bm E}\, 
{ K}({ \U}, \breve{\U})
~, 
\label{6.16}
\eea
where the dynamical variables $\U^I$ and their smile-conjugates $\breve{\U}^{\bar I}$ have the 
form 
\bea
{ \U}^I (\z) &=& \sum_{n=0}^{\infty}  \, \z^n \U_n^I  = \F^I + \z \S^I + \dots~, 
\qquad
\breve{ \U}^{\bar I} (\z) =\sum_{n=0}^{\infty}  \,  (-\z)^{-n}\,
{\bar \U}_n^{\bar I}~.
\label{6.17}
\eea
Here $\F^I:=\U^I_0 $ and $\S^I:=\U^I_1 $ are covariantly chiral and complex linear superfields, 
respectively, 
\bea
\bar{\bf D}_\a \F^I=0~, \qquad \bar{\bf D}^2 \S^I =0~, 
\eea
while the other components $\U^I_2, \U^I_3, \dots$, are unconstrained complex $\cN=2$ 
superfields. 

It is known that (2,0) AdS supersymmetry allows only $R$-invariant $\s$-model couplings
\cite{KT-M-2011}. 
As concerns the (2,1) supersymmetric $\s$-model \eqref{6.16}, it possesses the following U(1) 
symmetry: 
\bea
\U(\z) ~\to ~\U(\re^{\ri \a} \z) ~, \qquad \a \in {\mathbb R}~,
\eea
compare with \cite{GK1}. This symmetry is a special case of the transformation law
\eqref{5.35} obtained by setting $t = \a = \rm const $ and switching off the other parameters.

Upon reduction  to the (1,1) AdS superspace, 
the  supersymmetric $\s$-model \eqref{sigma-2_1} is described by the action
\bea
S =\hf  \oint_\g   \frac{\rd \zeta}{2\pi \ri \zeta}
 \int \rd^3x\, \rd^2\q \rd^2{\qb}\, {\cE}\, w^{ [2]} \,
 K({ \U}, \breve{\U})~, 
\qquad w^{[2]} = - \ri \Big( \frac{\bar w}{\z}+ w \z\Big)~.
\eea
The dynamical variables $\U^I (\z)$ and $\breve{\U}^{\bar I}(\z) $ have the functional form
\eqref{6.17} where $\F^I$ and $\S^I$ obey the constraints
\bea
\bar \nabla_\a \F^I = 0 ~, \qquad ( \bar \nabla^2 - 4\m ) \S^I =0~,
\eea
and the other components $\U^I_2, \U^I_3, \dots$, are unconstrained complex $\cN=2$ 
superfields.

%%%%%%%%%%%%%%%%%%%%%%%%%%%%%%%%%%%%%%%%%%%%%%%
%%%%%%%%%%%%%%%%%%%%%%%%%%%%%%%%%%%%%%%%%%%%%%%
%%%%%%%%%%%%%%%%%%%%%%%%%%%%%%%%%%%%%%%%%%%%%%%

\section{Conclusion}
\setcounter{equation}{0}

In conclusion, we briefly summarise our main results and list some open problems.
In this paper we introduced the three-dimensional $(p,q)$ AdS superspaces, studied 
their geometric properties  
and proved their conformal flatness when $X^{IJKL}=0$. Building on the results of
\cite{KLT-M-2011}, we then developed the fully-fledged projective-superspace formalism 
to construct off-shell $\cN=3$ rigid supersymmetric field theories in $\rm AdS_3$. 
There are two types of such theories, with (3,0) and (2,1) AdS supersymmetry respectively. 
We are especially interested in theories possessing $(p,q)$ AdS supersymmetry with 
$\cN= p+q\leq4$ because {\it nonlinear} $\s$-models exist only in these cases. 
We recall that the $\s$-models with $\cN= p+q =2$ were studied earlier 
in \cite{IT,DKSS,KT-M-2011}. 
The explicit construction of $(p,q)$ supersymmetric 
$\s$-models with $p+q=3$ was the subject of the present work. 
An open interesting problem is to extend our analysis given in this paper to the cases 
$\cN= p+q=4$. Conceptually, this should be similar to the $\cN=3$ case studied above, 
however some nontrivial new aspects will emerge. 
In particular, of special interest are those (4,0) supersymmetric $\s$-models 
which correspond to the extremal case \eqref{4.21}.

In this paper we constructed the general (3,0) and (2,1) supersymmetric $\s$-models
described by off-shell polar  hypermultiplets defined on the (3,0) and (2,1) 
AdS superspaces respectively. We then reduced these $\s$-models to
certain $\cN=2$ AdS superspaces. An interesting open problem is to reformulate the $\s$-models 
obtained in terms of $\cN=2$ chiral superfields in $\rm AdS_3$. (The importance of such a 
formulation is that  
it should  provide a direct access to the hyperk\"ahler geometry 
of the target space \cite{HKLR,BKsigma1,BKsigma2}.)    
This can be achieved by generalising the approaches developed in 
\cite{K-duality,K-comments,BKLT-M}.

As shown in this paper, the $\cN=3$ AdS supersymmetry imposes nontrivial restrictions 
on the $\s$-model hyperk\"ahler target spaces. The most unexpected outcome is that (3,0) 
AdS supersymmetry requires the $\s$-model target spaces to be hyperk\"ahler cones. 
Nevertheless, this result has a natural geometric origin. 
The main difference between the two types of $\cN=3$ supersymmetric $\s$-models 
in $\rm AdS_3$ is encoded in the corresponding  $R$-symmetry groups: SO(3) in the (3,0) case 
and SO(2) in the (2,1) case. It can be shown that any one-dimensional 
subgroup $H = \rm SO(2) $ of the $R$-symmetry group 
acts faithfully by rotations on the two-sphere of complex structures of the hyperk\"ahler target 
space $(\cM , g, \cJ_A )$.\footnote{The existence of such hyperk\"ahler spaces was pointed out
twenty five years ago in \cite{HitchinKLR}.} 
Here $g_{\m\n} $ is the hyperk\"ahler metric, and 
 $(\cJ_A)^\m{}_\n$ is the complex structures of $\cM$, 
$\cJ_A = (\cJ_1, \cJ_2, \cJ_3)$,  obeying  the quaternionic algebra
$\cJ_A \cJ_B = -\delta_{A B} {\mathbbm 1} + \ve_{ABC} \cJ_C.$
Suppose that $\cJ_3$ is invariant under the action of the subgroup $H$, 
and let $V^\m$ be  the Killing vector $V^\m$ associated with $H$. 
Without loss of generality, we have
\begin{align}\label{eq_VrotatescJ}
\cL_V \cJ_1 = - \cJ_2~, \quad
\cL_V \cJ_2 = + \cJ_1~, \quad
\cL_V \cJ_3 = 0~.
\end{align}
The Killing vector $V^\m$ 
 is holomorphic with respect to $\cJ_3$, and
we can introduce the corresponding Killing potential $\cK$ defined by 
\bea
V^\m = \hf (\cJ_3)^\m{}_\n \nabla^\n \cK~.
\eea 
As shown in \cite{BKLT-M},  $\cK$ is a globally defined function over $\cM$, and is
the K\"ahler potential with respect to $\cJ_1$ and $\cJ_2$ and indeed
any complex structure $\cJ_\perp$ which is perpendicular to $\cJ_3$. In other words,
\begin{align}
g_{\mu \nu} &= \frac{1}{2} \nabla_\mu \nabla_\nu \cK
	+ \frac{1}{2} (\cJ_\perp)_\mu{}^\rho (\cJ_\perp)_\nu{}^\sigma \nabla_\rho \nabla_\sigma \cK~.
\end{align}
It follows that the K\"ahler forms associated with $\cJ_1$ and $\cJ_2$ are exact, 
and thus $\cM$ is non-compact \cite{BKsigma1,BKsigma2,BKLT-M}.
As shown in \cite{BKLT-M}, the K\"ahler potential $\mathbb K$ with respect to $\cJ_3$ 
can be chosen such that 
\bea
(\cJ_3)^\m{}_\n V^\n \,{\mathbb K}_\m= - \cK~. 
\eea
So far, we have taken into account only the fact that the $R$-symmetry group contain 
a subgroup SO(2). In the case that the $R$-symmetry group coincides with SO(3), 
the above consideration implies that ${\mathbb K} = \cK$, 
and hence 
\bea
\nabla^\m \cK \nabla_\m\cK= 2 \cK~. 
\eea
We further deduce that $\c^\m = g^{\m \n} \nabla_\n \cK$ is a homothetic conformal Killing vector, 
\bea 
\nabla_\n \c^\m = \d^\m_\n~, 
\eea
and therefore $\cM$ is a hyperk\"ahler cone \cite{GR,deWRV}. 
In regard to the above discussion, we should also mention an interesting work \cite{IN} 
in which it was shown that  a sufficient condition for a
4D $\cN=2$ $\s$-model in projective superspace 
to be  superconformal is that its  $R$-symmetry is SO(3). 

The supergravity techniques of \cite{KLT-M-2011} can straightforwardly be applied to construct off-shell 
$\s$-models in the deformed $\cN=4$ Minkowski superspace described by covariant derivatives
obeying the (anti) commutation relations 
\bsubeq
\bea
\{\cD_\a^{i\bai},\cD_\b^{j\baj }\}&=&
2\ri\ve^{ij}\ve^{\bai \baj }\cD_{\a\b}
+{2\ri}\ve_{\a\b}\ve^{\bai \baj }X\bL^{ij}
-2\ri\ve_{\a\b}\ve^{ij}X \bR^{\bai \baj }
~,
\\
%%%%%%%%%%%%
{[}\cD_a,\cD_\b^{j\baj}{]}&=&0
~,~~~~~~
{[}\cD_a,\cD_b{]}=0
\eea
\esubeq
which follow from \eqref{4200} by setting $S=0$. 
An interesting open problem is to understand the target space geometry of such 
$\cN=4$ supersymmetric nonlinear $\s$-models. 
\\

\noindent
{\bf Acknowledgements:}\\
SMK is grateful to Daniel Butter for useful discussions. 
We are grateful to the program {\it Geometry of Strings and Fields} at Nordita, Stockholm
 (November, 2011)
where part of this work was
carried out, for providing a stimulating atmosphere.
GT-M thanks the School of Physics at the University of Western Australia
and the Theory Unit at CERN for the kind hospitality and support. He 
also thanks the Departments of Physics of Milano-Bicocca University and Milano University 
for kind hospitality during various stages of this work.
The work of SMK  is supported in part by the Australian Research Council. 
The work of UL was supported by VR-grant 621-2009-4066.
The work of GT-M was supported by the European Commission, Marie Curie Intra-European 
Fellowships under contract No. PIEF-GA-2009-236454. 
GT-M is the recipient of an Australian Research Council's Discovery Early Career Award 
(DECRA), project No. DE120101498.

%%%%%%%%%%%%%%%%%%%%%%%%%%%%%%%%%%%%%%%%%%%%%%
%%%%%%%%%%%%%%%%%%%%%%%%%%%%%%%%%%%%%%%%%%%%%%
%%%%%%%%%%%%%%%%%%%%%%%%%%%%%%%%%%%%%%%%%%%%%%

\appendix
\section{Derivation of (\ref{components-Ac})}

Here we sketch the derivation of the action (\ref{components-Ac}) 
by requiring its invariance under the projective transformations (\ref{proj-inv}).
The derivation is actually similar to those given in  \cite{KT-M,KT-M-normal,KT-M_5D},
and the interested reader is referred to those papers for more technical details.  

The strategy is to start from the zero-order term
$S_0$ in (\ref{comp-Ac-000}), vary it under the infinitesimal transformation (\ref{proj-inv})
 and add  iteratively extra terms to the action, which cancel the variation order by order, 
 such that the final action is invariant.
Instead of working with the general infinitesimal transformation (\ref{proj-inv}), it suffices to deal 
with the $b$ variation
\be
\d u_i=\,b\,v_i~,
\label{dubv}
\ee
since 
the $a$ and $c$ variations do not contribute if 
 degrees of homogeneity in $v$ and $u$ are chosen properly.
The transformation (\ref{dubv}) induces the following variations: 
\bsubeq
\bea
&&~~~~~~~~~~~~~~~
\d\cD_\a^{(-2)}=\frac{2b}{(v,u)}\cD_\a^{(0)}~,~~~
\d\cD_\a^{(0)}=\frac{b}{(v,u)}\cD_\a^{(2)}~,~~~
\label{d1}
\\
&&
\d\cS^{(-4)}=\frac{4b}{(v,u)}\cS^{(-2)}~,~~~
\d\cS^{(-2)}=\frac{3b}{(v,u)}\cS^{(0)}~,~~~
\d\cS^{(0)}=\frac{2b}{(v,u)}\cS^{(2)}~,
\label{d2}
\eea
\esubeq
where $\cS^{(2)}:=(v_iv_jv_ku_l\cS^{ijkl})/{(v,u)}$.
Let us compute the variation of $S_0$ defined by  (\ref{comp-Ac-000}).
Making use of (\ref{d1})--(\ref{d2}) and the analyticity condition $\cD_\a^{(2)}\cL^{(2)}=0$
gives
\bea
\d S_0&=&
\frac{1}{8\p}\int {\rm d}^3x\, e 
 \oint_{\g}  { v_i {\rm d} v^i }\frac{b}{(v,u)}\Big{[}
2\{\cD^{(0)\a},\cD^{(-2)}_\a\} \big(\cD^{(0)}\big)^2 
+4\cD^{(-2)\a} \cD^{(0)}_\a \big(\cD^{(0)}\big)^2 
\non\\
&&
+   \big(\cD^{(-2)}\big)^2 \{\cD^{(2)\a},\cD^{(0)}_\a\}
\Big{]}\cL^{(2)}||~.
\label{var-1}
\eea
The integrand can be considerably simplified. 
Using the algebra of covariant derivatives, \eqref{alg-3331}--\eqref{alg-3332},
it is not difficult to derive the following relation
\bea
 \cD^{(0)}_\a \big(\cD^{(0)}\big)^2\cL^{(2)} =
\Big(
\ri \cD_{\a\b}\cD^{(0)\b}
+\ri(2\cS^{(0)}-\cS)\cD^{(0)}_\a
+\ri\cS^{(2)}\cD_{\a}^{(-2)}
\Big)\cL^{(2)}
~,
\label{us1}
\eea
which has to be plugged in eq. (\ref{var-1}).
Next, we evaluate the anti-commutators in (\ref{var-1}) and
iteratively move all the
Lorentz and SU(2) generators to the right. Once they hit $\cL^{(2)}$ we use the identities
$\cM_{\a\b}\cL^{(2)}=v_iv_j\cJ^{ij}\cL^{(2)}=0$ and $v_iu_j\cJ^{ij}\cL^{(2)}=-(v,u)\cL^{(2)}$.
To compute the contributions coming from $u_iu_j\cJ^{ij}\cL^{(2)}$
one has to use the following formula 
\bea
\oint {v_i\rd{v}^i\over (v,u)^6}\,b\, \cT^{(3)}u_iu_j\cJ^{ij} \cL^{(2)}=
\oint  {v_i\rd{v}^i\over (v,u)^5}
\Big\{
b\Big(u^k\frac{\pa}{\pa v^k}\cT^{(3)}\Big) \cL^{(2)}
\Big\}
~.
\eea
This can be obtained using the results of \cite{KT-M_5D,KT-M-normal},
and it holds for any operator $\cT^{(3)}$ which is a function of $v$ and $u$ and 
homogeneous in $v$ of degree three : $\cT^{(3)}(cv)=c^3\cT^{(3)}(v)$.
The next step is to 
simplify the expression  (\ref{var-1}) obtained 
by moving the  vector derivative $\cD_{\a\b}$
coming from (\ref{us1})
to the left, which gives a total derivative to be ignored.
The final result is
\bea
\d S_0&=&
\frac{1}{8\p}\int {\rm d}^3x\, e 
 \oint_{\g}  { v_i {\rm d} v^i }\frac{b}{(v,u)}\Big{[}
\,16\ri\cS^{(2)}(\cD^{(-2)})^2
-4\ri(\cS^{(0)}+4\cS)\cD^{(-2)\a}\cD^{(0)}_\a
\non\\
&&~~~~~~~~~~~~~~~~~~~~~~~~~~~~~~~~~~
+40\ri\cS^{(-2)}(\cD^{(0)})^2 
+96\cS^{(2)}\cS^{(-4)}
    \Big{]}\cL^{(2)}||~.
\eea
To cancel this variation, we  consider an additional functional of the form
\bea
S_{\rm extra}&=&
\int {\rm d}^3x\, e 
 \oint_{\g} \frac{ { v_i {\rm d} v^i }}{8\p}
\Big{[}
\, \ri(a_1\cS^{(0)}+a_2\cS)(\cD^{(-2)})^2
 +a_3 \ri\cS^{(-2)}\cD^{(-2)\a}\cD_\a^{(0)}
+ a_4 \ri\cS^{(-4)}(\cD^{(0)})^2
 \non\\
 &&~~~~~~~~~~~~~~~~~~~~~~~
+a_5 \cS^{(-2)}\cS^{(-2)}
+a_6 \cS^{(-4)}\cS^{(0)}
+a_7 \cS^{(-4)}\cS
\,\Big{]} \cL^{(2)}||~.
\eea
By using the procedure described for the computation of $\d S_0$, we derive
\bea
\d S_{\rm extra}&=&
\frac{1}{8\p}\int {\rm d}^3x\, e 
 \oint_{\g}  { v_i {\rm d} v^i }
 \frac{b}{(v,u)}
\Big{[}\,
 2\ri a_1\cS^{(2)}(\cD^{(-2)})^2
+\,2\ri(a_3+2a_4)\cS^{(-2)}(\cD^{(0)})^2
\non\\
&&
+  \ri\Big((4a_1+3a_3)\cS^{(0)}+4a_2\cS\Big)\cD^{(-2)\a}\cD_\a^{(0)}
+\big(-12a_4+2a_6\big)\cS^{(-4)}\cS^{(2)}
\non\\
&&
+\,\cS^{(-2)} \Big((-24a_1+6a_5+4a_6)\cS^{(0)}
+(16a_1-16a_2+4a_7)\cS\Big)
\Big{]} \cL^{(2)}||
~.
~~~~~~
\eea
Imposing the condition  $\d S_0+\d S_{\rm extra}=0$ fixes the coefficients  
\bea
a_1=-8~,~~
a_2=4~,~~
a_3=12~,~~
a_4=-16~,~~
a_5=-144~,~~
a_6=64~,~~
a_7=48~.~~~~~~
\eea
These results give the desired action (\ref{components-Ac}).

%%%%%%%%%%%%%%%%%%%%%%%%%%%%%%%%%%%%%%%%%%%%%%%%
%%%%%%%%%%%%%%%%%%%%%%%%%%%%%%%%%%%%%%%%%%%%%%%%
%%%%%%%%%%%%%%%%%%%%%%%%%%%%%%%%%%%%%%%%%%%%%%%%

\begin{footnotesize}

\end{footnotesize}

%%%%%%%%%%%%%%%%%%%%%%%%%%%%%%%%%%%%%%%%%%%%%%
%%%%%%%%%%%%%%%%%%%%%%%%%%%%%%%%%%%%%%%%%%%%%%
%%%%%%%%%%%%%%%%%%%%%%%%%%%%%%%%%%%%%%%%%%%%%%

%%%%%%%%%%%%%%%%
%%%%%%%%%%%%%%%%
\end{document}